\documentstyle[12pt]{article}

\begin{document}
\title{\bf Correlations within Eigenvectors and Transition
Amplitudes in the Two-Body Random Interaction Model}

\author{V.V.Flambaum,$^{1,2}$\thanks{email:
flambaum@newt.phys.unsw.edu.au}~~G.F.Gribakin,$^{1}$\thanks{email:
gribakin@newt.phys.unsw.edu.au}~~and
F.M.Izrailev$^{1,2}$\thanks{email: izrailev@physics.spa.umn.edu}\\
\\
\small $^1$ School of Physics, University of New South Wales,
\small Sydney 2052, Australia\\
\small $^2$ Budker Institute of Nuclear Physics, 630090 Novosibirsk,
Russia }

\maketitle
\begin{abstract}
It is shown that the two-body character of the interaction in a
many-body system gives rise to specific correlations between the
components of compound states, even if this interaction is completely
random. Surprisingly, these correlations increase with the increase of
the number of active (valence) particles. Statistical theory of
transition amplitudes between compound states, which takes into
account these correlation is developed and tested within the framework
of the Two-Body Random Interaction Model. It is demonstrated that a
feature, which can be called ``correlation resonance'', appears in the
distribution of the transition matrix amplitudes, since the
correlations strongly reduce the transition amplitudes at the tails
and increase them near the maximum of the distribution.
\end{abstract}

\newpage
\sloppy

\section{Introduction}

Measurements of parity non-conservation in neutron capture by Th
nucleus \cite{FR91} gave a surprising result: in spite of a natural
assumption of a random character of matrix elements of the weak
interaction between compound states, the effect was found to be of the
same sign for all observed resonances. Possibly, this means that the
strongly fluctuating matrix elements of a weak perturbation between
``chaotic'' states of a compound nucleus (which was, in fact, the
first example of quantum chaotic system) are essentially correlated.
Thus, the problem of correlations between components of compound
states is of great importance both for theory and applications.

An attempt to study these correlations has been undertaken in
\cite{F92,F94,FG95}. It is natural to expect that some correlations
may appear if the number of independent parameters in a Hamiltonian
matrix is substantially smaller that the total number of the
Hamiltonian matrix elements. The simplest example is given by the
model of a random separable interaction \cite{F95} (see also
\cite{FG95}),
\begin{equation}
\label{Hsep}H_{ij}=\epsilon _i\delta _{ij}+gv_iv_j,\quad i,j=1,2,
\dots ,N,
\end{equation}
where $\epsilon _i$ are the unperturbed energies and $v_i$ are random
variables distributed, e.g., according to a Gaussian law. As one can
see, the number of independent parameters $v_i$ is equal to $N$, while
the number of the Hamiltonian matrix elements $H_{ik}$ is $N^2$. It
was shown that this model displays very strong ($\sim 100\%$)
correlations between eigenvectors with close energies, despite the
random character of the interaction: $\langle v_iv_j\rangle =
\delta _{ij}\langle v_i^2\rangle $.

Such correlations cannot appear in models described by full random
matrices, like those of the Gaussian Orthogonal Ensemble (GOE). It was
always obvious that such models possess some unphysical features,
e.g., the semicircle level density, however, they seem to give a very
accurate description of the correlations and fluctuations of energy
levels (see, e.g., \cite{brody}). The important point is that in real
many-body systems the basic interaction is a two-body one. This means
that the number of independent parameters determining the $n$-particle
Hamiltonian (two-body matrix elements) is much smaller than the number
of the Hamiltonian matrix elements. Taking the two-body matrix
elements as Gaussian random variables, a model called the Two-Body
Random Ensemble (TBRE) was introduced in \cite{FW70,BF71} (see also
references in \cite{brody}). This model looks in principle much more
realistic than the GOE. In particular, it has a Gaussian form of the
level density, which is in good agreement with various nuclear
shell-model calculations for ``realistic'' interactions in the finite
basis (see, e.g., \cite{zelev}). The TBRE does not allow a
deep analytical treatment, however, numerical modeling showed that
its level fluctuation properties are very close to those of the GOE,
although some differences were noticed \cite{BF71}. Therefore, in some
respect, the two-body nature of the particle interaction does not
reveal itself in the level statistics. In other words, level
fluctuations are insensitive to the details of the interaction
between particles, provided the latter is large
enough to cause strong mixing of the basis states.

Another model which seems to be more physical then the GOE was
proposed a while ago by Wigner \cite{wigner} to describe compound
nuclei. It was suggested that the Hamiltonian matrix $H_{ij}$ has a
banded structure, i.e., all matrix elements with $|i-j|>b$ are zeros
($b$ is the bandwidth). The matrix elements inside the band are random
numbers with zero mean and fixed variance, $\overline{H_{ij}}=0$,
$\overline{H_{ij}^2}=V^2$, except those on the main diagonal, which
monotonically increase, $H_{ii}=iD$. In this model all eigenstates are
localized in the unperturbed basis ($V=0$). In the non-perturbative
regime, $1<V/D<\sqrt{b}$ the strength function which describes the
localization in the energy space (also called the local spectral 
density of states), has a characteristic Breit-Wigner form
with a width $\Gamma =2\pi V^2/D$ within the band. This shape is in
agreement with nuclear data \cite{BM}, and with the calculated
localization properties of chaotic eigenstates in the rare-earth atom
of Ce
\cite{FGGK94}. It has also been shown in \cite{FGGK94,FGG95} that the
Hamiltonian matrix which produces the dense spectrum of atomic
excited states in Ce, is sparse and has a band-like structure,
although the edges of the band are diffuse. We should also mention
that a number of analytical results on the localization properties of
such Band Random Matrices (BRM) has been recently obtained in
\cite{FM,CFIC} (see also \cite{CCGI} and references therein). However,
the off-diagonal Hamiltonian matrix elements in this BRM model are
independent random variables, thus this model is void of any possible
correlations related to the two-body interaction between particles.

In this paper we show that there are quantities (transition
amplitudes, or transition strengths) for which the underlying two-body
interaction is of crucial importance. We show that the such
interaction gives rise to specific correlations between the
components of eigenstates, which are very essential for the
distribution of transition strengths. Our results are obtained in
the framework of the Two-Body Random Interaction Model (TBRIM)
recently proposed in \cite{FIC95} for the study of various physical
problems related to such complex many-body systems as heavy atoms,
nuclei, metallic clusters, etc., which display quantum chaotic
behaviour. Being in some aspects similar to the TBRE, the TBRIM is
simpler in the sense that it abandons all restrictions imposed by the
conservation of the angular momentum which makes it closer to the
embedded GOE \cite{MF75}. On the other hand, the non-degenerate
spectrum of the single-particle orbitals the TBRIM is based upon,
generates a realistic level density and leads to a band-like
structure of the Hamiltonian matrix.

We should mention that there was quite a number of earlier works where
strength distributions were studied using statistical spectroscopy
methods and nuclear shell-model calculations (\cite{DFW77}, see also
review \cite{brody}). These methods are based on the calculation of
distribution moments, which are given by traces of products of the
operators in question and powers of the Hamiltonian over the model
finite-dimensional space of the problem. Since the calculation of
traces does not require knowledge of eigenstates, the question of
correlations within eigenstates which is of prime
importance for the present work, has not been addressed in those
studies. We must add that statistical spectroscopy methods emphasize
and employ a particular importance of Gaussian spreading of
many-particle configurations, and features like Breit-Wigner
localization either do not appear, or are neglected (together with
the interaction between configurations) in that formalism. All in all,
it is unfortunately very difficult for the present approach to make
contact with those results. Comparing the two approaches we should say
that at first sight ours does not look as rigorous and mathematically
advanced as the other one, as it appeals to some heuristic arguments
and uses rather simple mathematics, e.g., perturbation theory.
However, we believe that supported by numerical experiments, our
method can give a deeper insight and a more physical picture of
transitions between and correlations within the chaotic eigenstates
in complex many-body systems.

In Section 2 of this paper we show how the basic two-body interaction
results in the correlations between the Hamiltonian matrix elements,
eigenstate components and transition amplitudes. In Sec. 3 we check,
whether the effects found in Sec. 2 could lead to some correlations
between transition amplitudes coupling different pairs of states.
Section 4 presents a brief outline of a statistical approach to the
calculation of transition strengths; the analytical results are
checked there against numerical ones obtained in the TBRIM. Finally,
in Sec. 5 we study the spreading widths of the many-particle basis
states.

\section{Correlations between eigenvector components induced by
two-body interaction}

Let us consider the basic ideas of the TBRIM. In this model, $n$
Fermi-particles are distributed among $m$ non-degenerate orbitals.
In doing numerical experiments, we assume, as in \cite{FIC95}, that
the energies of the orbitals are given by the simple expression: 
\begin{equation}
\label{orbit}\epsilon _\alpha =d_0\left( \alpha +\frac 1\alpha
\right)~, \quad \alpha =1,2,...,m~.
\end{equation}
However, the analytical treatment presented below does not depend on a
particular form of $\epsilon _\alpha $ . Many-particle basis states
$\left| i\right\rangle $ are constructed by specifying the $n$
occupied orbitals. The energy $E_i$ of the basis state equals the sum
of the single-particle energies over the occupied orbitals. The total
number of the many-particle states in the model is
$N=\frac{m!}{n!(m-n)!}\sim \exp (n\ln \frac mn+(m-n)\ln
\frac m{m-n})$. The latter estimate relates to large $m$ and $n$
and shows that $N$ is exponentially large for $n,m-n\gg 1$ .

The number of independent parameters of the many-body Hamiltonian is
given by the number of different two-body interaction matrix elements
$V_{\alpha \beta \gamma \delta }$ and equals $N_2=m^2(m-1)^2/2$ . Due
to the two-body character of the interaction,
the Hamiltonian matrix element $H_{ij}=
\left\langle i\right| H\left| j\right\rangle $ is non-zero only when
$\left| i\right\rangle $ and $\left| j\right\rangle $ differ by no
more than two occupied single-particle orbitals. As a result, the
number $K$ of the non-zero matrix elements $H_{ij}$ is given by
\begin{displaymath}
K=N(K_0+K_1+K_2)~,
\end{displaymath}
\begin{equation}
K_0=1~,\quad K_1=n(m-n),\quad K_2=\frac{1}{4}n(n-1)(m-n)(m-n-1)~,
\label{K123}
\end{equation}
where $K_0$, $K_1$ and $K_2$ are the numbers of the Hamiltonian matrix
elements coupling a particular basis state $i$ to another one, $j$,
which differs form $i$ by the positions of none, one and two
particles, respectively. Therefore, for $n,m-n\gg 1$ we have
$N_2\ll K\ll N^2$, i.e., the Hamiltonian
matrix is essentially sparse and, in a sense, strongly correlated.

To see the correlation between non-zero matrix elements,
let us consider a pair of basis states $\left| i\right\rangle $ and
$\left| j\right\rangle $ which differ by the states of two particles,
for example, the state $\left| j\right\rangle $ can be obtained from
the state $\left| i\right\rangle $ by transferring the particles from
the orbitals $\alpha ,\beta $ into the orbitals $\gamma ,\delta $. For
all such pairs, the Hamiltonian matrix elements are the same,
$H_{ij}=V_{\alpha \beta \gamma \delta }$ (or, strictly speaking,
$H_{ij}=\pm V_{\alpha \beta \gamma\delta }$, due to Fermi statistics).
It is easy to calculate the total number $N_{eq}$ of the matrix
elements $H_{ij}$ equal to $V_{\alpha \beta \gamma \delta }$, using
the fact that the remaining $n-2$ particles can be arbitrary
distributed over $m-4$ orbitals,
\begin{equation}
\label{Neq}N_{eq}=\frac{(m-4)!}{(n-2)!(m-n-2)!}~.
\end{equation}
For basis states $\left| i\right\rangle $ and $\left| j\right\rangle $
which differ by the state of one particle $(\alpha \rightarrow
\beta )$ the matrix element $H_{ij}$ equals the sum of the $n-1$
two-body interaction matrix elements, $H_{ij}=\sum_\gamma V_{\alpha
\gamma \beta \gamma }$ (the index $\gamma $ runs over the rest $n-1$
occupied orbitals). In this case $H_{ij}$ for different $\left| i
\right\rangle $ and $\left| j\right\rangle $ (with fixed $\alpha $
and $\beta $) do not coincide, but may contain identical terms
$V_{\alpha \gamma \beta \gamma }$ , i.e., they are also correlated.

The eigenstates $\left| n_1\right\rangle $ of the model are determined
by their components $C_i^{(n_1)}$ with respect to the many-particle
basis states $\left|i\right\rangle $,
\begin{equation}
\label{ev}\left| n_1\right\rangle =\sum_iC_i^{(n_1)}\left| i\right
\rangle ~,
\end{equation}
and can be found by solving the Schr\"odinger equation, 
\begin{equation}
\label{schrod}\sum_jH_{ij}C_j^{(n_1)}=E^{(n_1)}C_i^{(n_1)}~.
\end{equation}
If the perturbation $V$ is strong enough, the exact eigenstates
$|n_1\rangle $ are superpositions of a large number of basis states.
As is known, strong mixing of basis states
in the exact eigenstates (compound states) occurs locally within some
energy range, $\left| E_i-E_{}^{(n_1)}\right| \leq \Gamma ,$ where
$\Gamma $ is known as the ``spreading width''. It can be estimated as
$\Gamma \approx N_wD$, where $D$ is the local mean level spacing for
many-particle states and $N_w$ is effective number of basis states
represented in a compound state. This number is also known as the
number of ``principal components''. These components give the main
contribution to the normalization condition
$\sum_i\left| C_i^{(n_1)}\right| ^2=1$ for the eigenstate
$\left| n_1\right\rangle $. Formally, we can estimate $N_w$ as the
reciprocal of the inverse participation ratio,
$N_w^{-1}\simeq \sum _i\left| C_i^{(n_1)}\right| ^4$.

It is rather straightforward to show that the correlations between
$H_{ij}$ result in correlations between the components $C_i^{(n_1)}$.
Indeed, let us multiply the Schr\"odinger equation by the coefficient
$C_i^{(n_1)}$ and sum over $n_1$. Using the orthogonality condition
$\sum_{n_1}C_i^{(n_1)}C_j^{(n_1)}=\delta _{ij}$, one obtains 
\begin{equation}
\label{ortho}H_{ij}=\sum_{n_1}C_i^{(n_1)}E^{(n_1)}C_j^{(n_1)}~.
\end{equation}
In what follows, we assume that the matrix elements of the two-body
interaction $V$ are random variables with the zero mean, therefore,
$\overline{H_{ij}}=0\,$ for $i\neq j$. In this case one can get
$\overline{C_i^{(n_1)}C_k^{(n_1)}}=0$ where the line stands for
averaging over different realizations of $V$. However, if matrix
elements of the Hamiltonian are correlated, $\overline{H_{ij}H_{kl}}
\neq 0$, the components of different eigenvectors $\left| n_1\right
\rangle $ and $\left| n_2\right\rangle $ are also correlated, since 
\begin{equation}
\label{Hcorr}\overline{H_{ij}H_{kl}}=\overline{
\sum_{n_1n_2}C_i^{(n_1)}E^{(n_1)}C_j^{(n_1)}C_k^{(n_2)}E^{(n_2)}
C_l^{(n_2)}}\neq 0 ~.
\end{equation}
The latter relation shows that $\overline{C_i^{(n_1)}C_j^{(n_1)}
C_k^{(n_2)}C_l^{(n_2)}}\neq 0$.

The above conclusion has important consequences. Let us consider a
single-particle operator 
\begin{equation}\label{M}
\hat M=\sum_{\alpha \beta }a_\alpha ^\dagger a_\beta ^{}M_{\alpha
\beta }\,=^{}\sum_{\alpha \beta }\rho _{\alpha \beta }\,M_{\alpha
\beta } 
\end{equation}
where $a_\alpha ^\dagger $ and $a_\beta $ are the creation and
annihilation operators. It is convenient to express the matrix
elements of $\hat M$ in terms of matrix elements of the density matrix
operator $\rho _{\alpha \beta }=a_\alpha ^\dagger a_\beta $ which
transfers a particle from the orbital $\beta $ to the orbital
$\alpha$. One can see that the matrix element of
$\hat M$ between compound states,%
$$
\left\langle n_1\right| \hat M\left| n_2\right\rangle =\sum_{\alpha
\beta }M_{\alpha \beta }\left\langle n_1\right| \rho _{\alpha \beta }
\left|n_2\right\rangle = 
$$
\begin{equation}
\label{n1n2}=\sum_{\alpha \beta }M_{\alpha \beta
}\sum_{ij}C_i^{(n_1)}\left\langle i\right| \rho _{\alpha \beta }\left|
j\right\rangle C_j^{(n_2)} 
\end{equation}
has the zero mean due to the statistical properties of the components,
i.e., $\overline{\left\langle n_1\right| \rho _{\alpha \beta }\left|
n_2\right\rangle }=0$. Since the summation over the orbitals $\alpha ,
\beta $ in Eq.(\ref{n1n2}) is independent from the averaging over
different
realizations of $V$, in what follows we consider the simplest case of
$\hat M=\rho _{\alpha \beta }$. The variance of the matrix element of
$\rho _{\alpha \beta }$ between the two compound states is equal to 
\begin{eqnarray}
\overline{M^2}&=&\overline{\left\langle n_1\right| \rho _{\alpha
\beta }\left|
n_2\right\rangle \left\langle n_2\right| \rho _{\beta \alpha }\left|
n_1\right\rangle }\nonumber \\
&=&\overline{\sum
\limits_{ijkl}C_i^{(n_1)}C_j^{(n_1)}C_k^{(n_2)}C_l^{(n_2)}\left\langle
i\right| \rho _{\alpha \beta }\left| k\right\rangle \left\langle l
\right|
\rho _{\beta \alpha }\left| j\right\rangle }\nonumber \\
&=&S_d^{(n_1n_2)}+S_c^{(n_1n_2)}~,\label{M2} 
\end{eqnarray}
where we separated the diagonal and non-diagonal contributions to the
sum (\ref{M2}),
\begin{equation}
\label{Sd}S_d^{(n_1n_2)}=\overline{\sum\limits_{ik}\left| C_i^{(n_1)}
\right|
 ^2\left| C_k^{(n_2)}\right| ^2\left| \left\langle i\right|
\rho _{\alpha \beta }\left| k\right\rangle \right| ^2}~, 
\end{equation}
\begin{equation}
\label{Sc}S_c^{(n_1n_2)}=\overline{\sum\limits_{i\neq j,\,k\neq
l}C_i^{(n_1)}C_j^{(n_1)}C_k^{(n_2)}C_l^{(n_2)}\left\langle i\right|
\rho _{\alpha \beta }\left| k\right\rangle \left\langle l\right| \rho
_{\beta \alpha }\left| j\right\rangle }~.
\end{equation}
Note that the diagonal term $S_d^{(n_1n_2)}\,$ is essentially
positive and can be easily estimated (see \cite{F94,FGGK94,FV93} and
Sec. 4 below), while the
non-diagonal term $S_c^{(n_1n_2)}$ is of our main interest. If the
eigenstates are completely ``random'' (different components both
inside each eigenstate and of different eigenstates are uncorrelated),
the correlation sum $S_c$ is equal to zero and the variance is
determined by the ``diagonal'' sum $S_d$ (this assumption have been
used in the previous calculations of matrix elements between compound
states in \cite{F94,FGGK94,FV93,SF82}). However, we show below that
in many-body system these two terms are of the same order,
$S_c\sim S_d$, even for the random two-body interaction $V$ .

The TBRIM allows one to investigate various properties of chaotic
many-body systems taking into account the two-body nature of the
interaction between particles. In the previous papers
\cite{FGGK94,FIC95} there were indications that the diagonal
approximation is not completely accurate for
the computation of the variance of matrix elements of perturbation. In
order to study this effect in detail, we have performed numerical
experiments with TBRIM for the parameters corresponding to the model
calculations of the Ce atom \cite{FGGK94,FGG95}. We take the number
of particles $n=4$, the number of orbitals $m=11$, the spectrum of the
single-particle orbitals is
determined by $d_0=1$, and the Gaussian random two-body interaction is
given by $\sqrt{\overline{V^2}}=0.12$. As a result, the size of the
Hamiltonian matrix $H_{ij}$ is $N=330$. The calculation of the matrix
elements between compound states in this model gave a remarkable
result. In Fig. 1 we present the ``experimental'' value of
$\overline{M^2}$ (see Eq. (\ref{M2})) together with the diagonal
contribution (\ref{Sd}) (see Fig. 1a), and the ratio $R=S_c/S_d$ in
Fig. 1b. Figure 1a reveals a systematic difference between the
diagonal approximation and exact expression (\ref{M2}), and Fig. 1b
shows that non-diagonal term $S_c$ is of the same order as $S_d$
which clearly indicates the presence of correlations. 

Below, we show how these correlations emerge in the non-diagonal term
$S_c$. First, note that for a given $i$ the sum over $k$ in Eq.
(\ref{Sd}) for $S_d$ contains only one term, for which $|k\rangle =
a_\beta ^\dagger a_\alpha |i\rangle \equiv |i'\rangle $, determined
by transferring one particle from the orbital $\alpha $ to the
orbital $\beta $ in the state $|i\rangle $ (hereafter we will use the
notation $i'$ to denote such states). Accordingly, the index $i$ runs
over those states in which $\alpha $ is occupied and
$\beta $ is vacant. For such $i$ and $i'$ the matrix element
$\left\langle i\right| \rho _{\alpha \beta }\left|i'\right\rangle \,
=1$, otherwise, it is zero. Therefore, in fact, the sum in (\ref{Sd})
is a single sum, with a number of items less than $N$,
\begin{equation}
\label{SSd}S_d^{(n_1n_2)}=\overline{{\sum_i}^\prime \left|
C_i^{(n_1)}\right| ^2\left| C_{i'}^{(n_2)}\right| ^2}
\end{equation}
where the sum ${\sum_i}^\prime $ runs over the specified $i$.
Analogously, Eq. (\ref{Sc}) can be written as the double sum over $i$
and $j$ specified as above,
\begin{equation}
\label{SSc}S_c^{(n_1n_2)}=\overline{{\sum\limits_{i\neq j}}^{\prime
\prime } C_i^{(n_1)}C_j^{(n_1)}C_{i'}^{(n_2)}C_{j'}^{(n_2)}},
\end{equation}
where $j'$ is a function of $j$, $|j'\rangle =a_\beta ^\dagger
a_\alpha |j\rangle $. Note that the energies of the basis states and
their primed partners are connected as $E_{i'}-E_i=\epsilon _\beta -
\epsilon _\alpha = E_{j'}-E_j$.

One can expect that maximal values of the sum (\ref{SSd}) and,
possibly, (\ref{SSc}) are achieved when $C$'s are principal components
of the eigenstates. This means that  mean square of the matrix element
$\overline{\left| \left\langle n_1\right| \rho
_{\alpha \beta }\left| n_2\right\rangle \right| ^2}$ is maximal when
the operator $\rho _{\alpha \beta }$ couples the principal components
of the state $\left| n_1\right\rangle $ with those of $\left| n_2
\right\rangle $, i.e. for $E^{(n_1)}-E^{(n_2)}\approx \omega _{\alpha
\beta }$ $\equiv \epsilon _\alpha -\epsilon _\beta $. Far from the
maximum ($\left| E^{(n_1)}-E^{(n_2)}-\omega _{\alpha \beta }\right|
>\Gamma $) a principal component of one state, say, $n_1$, is coupled
to a small component $k$ of the other state $n_2$ ($\left| E_k-
E^{(n_2)}\right| >\Gamma )$. The latter case is simpler to consider
analytically,  since the admixture of a small component in the
eigenstate can be found by means of perturbation theory. This approach
reveals the origin of the correlations in the sum $S_c$,
Eq.~(\ref{SSc}). For example, if $C^{(n_1)}_j$ is a small component of
the eigenstate $n_1$, then it can be expressed as a perturbation
theory admixture to the principle components. If $C^{(n_1)}_i$ is one
of the latter, then there is a term in the sum (\ref{SSc}), which is
proportional to the principal component squared, $\left|C^{(n_1)}_i
\right| ^2$.

Indeed, there are three possibilities:

(i) $C_i^{(n_1)}$ and $C_{j'}^{(n_2)}$ are among the principal
components, and $C_j^{(n_1)}$ and $C_{i'}^{(n_2)}$ correspond to the
small components. Then, one can write
\begin{equation}\label{C1}
C_j^{(n_1)}=\frac{\left\langle j\right| H\left| \tilde
n_1\right\rangle }{E^{(n_1)}-E_j}=\widetilde{\sum\limits_{p}}
\frac{H_{jp}}{E^{(n_1)}-E_j}C_{p}^{(n_1)}~,
\end{equation}
\begin{equation}\label{C2}
C_{i'}^{(n_2)}=\frac{\left\langle i'\right| H\left| \tilde
n_2\right\rangle }{E^{(n_2)}-E_{i'}}=\widetilde{\sum\limits_{q}}
\frac{H_{i'q}}{E^{(n_2)}-E_{i'}}C_q^{(n_2)}~.
\end{equation}
Tilde above the sums indicates that the summations runs over the
principal components only. The ``coherent'' contribution to the sum
$S_c$ in Eq.
(\ref{SSc}) is obtained by separating the squared contributions of the
principal components in the sums in $S_c^{(n_1n_2)}$ (i.e.,
$p=i,~q=j'$)
\begin{equation}\label{Sc1a}
\widetilde{{\sum\limits_{i,j'}}''}\frac{\overline{H_{i'j'}H_{ji}}}
{(E^{(n_2)}-E_{i'})(E^{(n_1)}-E_j)}\left| C_i^{(n_1)}\right| ^2
\left| C_{j'}^{(n_2)}\right|^2 ~.
\end{equation}
Taking into account that for the principal components we have
$E_i\approx E^{(n_1)}$ and $E_{j'}\approx E^{(n_2)}$, we can replace
the energies, $E_{i'}\rightarrow E^{(n_1)}+\omega _{\beta \alpha }$,
and $E_j\rightarrow E^{(n_2)}-\omega _{\beta \alpha }$, and thus
obtain the following contribution to $S_c^{(n_1n_2)}$:
\begin{equation}\label{Sc1b}
-\frac 1{(E^{(n_2)}-E^{(n_1)}-\omega _{\beta \alpha })^2} 
\widetilde{{\sum\limits_{i,j'}}''}\left| C_i^{(1)}\right| ^2\left|
C_{j'}^{(2)}\right| ^2\overline{H_{i'j'}H_{ij}}~.
\end{equation}

(ii) $C_j^{(n_1)}$ and $C_{i'}^{(n_2)}$ correspond to the principal
components, $C_i^{(n_1)}$ and $C_{j'}^{(n_2)}$ correspond to
the small components. Then, the result is the same as (\ref{Sc1b}).

(iii) $C_i^{(n_1)}$ and $C_j^{(n_1)}$ are principal components,
$C_{i'}^{(n_2)}$ and $C_{j'}^{(n_2)}$ are small components
(or, $C_{i'}^{(n_2)}$ and $C_{j'}^{(n_2)}$ are principal components,
$C_i^{(n_1)}$ and $C_j^{(n_1)}$ are small components). In these cases
there are no coherent terms in the sum for $S_c$ in Eq. (\ref{Sc}).
This follows from the fact that for chaotic eigenstates the mixing
among the principal components is practically complete, which makes
them to a good accuracy statistically independent.

Thus, far from the maximum, $\left| E^{(n_2)}-E^{(n_1)}-
\omega _{\beta \alpha }\right| >\Gamma $, one obtains 
\begin{equation}\label{Sc34}
S_c^{(n_1n_2)}\approx -\frac 2{\left( E^{(n_2)}-E^{(n_1)}-
\omega _{\beta \alpha }\right) ^2}\widetilde{{\sum\limits_{i,j'}}''}
\left| C_i^{(1)}\right| ^2\left|C_{j'}^{(2)}\right| ^2
\overline{H_{i'j'}H_{ij}}
\end{equation}
A similar calculation of the diagonal sum $S_d^{(n_1n_2)}$, Eq.
(\ref{Sd}), yields
\begin{eqnarray}
S_d^{(n_1n_2)}\!\!\! &\approx &\!\!\!
\frac 1{\left( E^{(n_2)}-E^{(n_1)}-\omega _{\beta \alpha}\right) ^2}
\nonumber \\
& \times &\!\!\!\! \left[ \widetilde{{\sum\limits_{i}}'}
\widetilde{{\sum\limits_{j'}}}\left| C_i^{(n_1)}\right| ^2
\left| C_{j'}^{(n_2)}\right| ^2 \overline{H_{i'j'}^2} +
\widetilde{{\sum\limits_{i}}}\widetilde{{\sum\limits_{j'}}'}
\left| C_i^{(n_1)}\right| ^2 \left| C_{j'}^{(n_2)}\right| ^2
\overline{H_{ij}^2} \right] .\label{Sd34}
\end{eqnarray}
The two terms in square brackets result from the contribution of
principal $i$ and small $i'$ components in Eq. (\ref{SSd}), and
{\em vice versa}. From Eq. (\ref{Sc34}) we see that $S_c^{(n_1n_2)}=0$
if $\overline{H_{i'j'}H_{ij}}=0$. However, there is nearly a hundred
percent correlation between these matrix elements. Indeed, the basis
state $i'$ differs from $i$ by the location of only one particle (the
transition from the orbital $\alpha $ to $\beta $), and the same is
true for $j'$ and $j$.

Let us estimate the relative magnitudes of $S_d$ and $S_c$.
First, consider the case when $|i\rangle $ and $|j\rangle $ differ by
two orbitals,  $|j\rangle =a^\dagger _{\mu _2}a _{\mu _1}
a^\dagger _{\nu _2}a_{\nu _1}|i\rangle $. In this case $H_{ij}=
V_{\nu _1\mu _1\nu _2\mu _2}$.
Since the basis states $|i'\rangle $ and $|j'\rangle $ must differ by
the same two orbitals, we have $H_{i'j'}=V_{\nu _1\mu _1\nu _2\mu _2}
=H_{ij}$ (note that $\nu _1,\mu _1,\nu _2,\mu _2\neq \alpha ,\beta $,
since both states $|i\rangle $ and $|j\rangle $ contain $\alpha $ and
do not contain $\beta $, whereas $|i'\rangle $ and $|j'\rangle $
contain $\beta $ and do not contain $\alpha $). Therefore, the
averages over the non-zero matrix elements
between such pairs of states are: $\overline {H_{ij}H_{i'j'}}=
\overline {H_{ij}^2}=\overline{H_{i'j'}^2}=V^2$.

Now, let us consider the case when $|i\rangle $ and $|j\rangle $
differ by one orbital $|j\rangle =a^\dagger _{\nu _2}a_{\nu _1}|i
\rangle $. In this case the Hamiltonian matrix elements are sums of
the $n-1$ two-body matrix elements,
\begin{eqnarray}
H_{ij}=\sum\limits_{\mu \neq \alpha }^{n-2}V_{\nu_1\mu \nu_2\mu }
+V_{\nu_1\alpha \nu_2\alpha }~,\nonumber \\
H_{i'j'}=\sum\limits_{\mu \neq \beta }^{n-2}V_{\nu_1\mu \nu_2\mu }
+V_{\nu_1\beta \nu_2\beta }~.\nonumber
\end{eqnarray}
The sums of $n-2$ terms in $H_{ij}$ and $H_{i'j'}$ coincide, the
difference is due to the one term only (orbital $\alpha $ is replaced
by the orbital
$\beta $). Thus,
\begin{eqnarray}
\overline{H_{ij}H_{i'j'}}=(n-2)V^2~,\nonumber \\
\overline{\left( H_{ij}\right) ^2}= 
\overline{\left( H_{i'j'}\right) ^2}=(n-1)V^2 ~,\nonumber
\end{eqnarray}
where we took into account that $\overline{V_{\kappa \lambda \mu \nu }
V_{\kappa _1\lambda _1\mu _1\nu _1}}=V^2\delta _{\kappa \kappa _1}
\delta _{\lambda \lambda _1}\delta _{\mu \mu_1}\delta _{\nu \nu _1}$.

The contributions of one-particle and two-particle transitions in Eqs.
(\ref{Sc34}) and (\ref{Sd34}) representing $S_c$ and $S_d$
respectively,
will be determined by the numbers of such transitions allowed by the
corresponding sums. For the single-prime sums in Eq. (\ref{Sd34})
these numbers are proportional to $K_1$ and $K_2$, Eq. (\ref{K123}).
In the double-prime sum in Eq. (\ref{Sc34}) these numbers are
proportional to $\tilde K_1$ and $\tilde K_2$, the numbers of the
two-body and one-body transitions $i\rightarrow j$, in the situation
when one particle and the two orbitals ($\alpha $ and $\beta $) do
not participate in the
transitions. These numbers can be obtained from Eq. (\ref{K123}) if we
replace $n$ by $n-1$, and $m$ by $m-2$, so that
$\tilde K_1=(n-1)(m-n-1)$, $\tilde K_2=(n-1)(n-2)(m-n-1)(m-n-2)/4$.
Finally we obtain that at $|E^{(n_2)}-E^{(n_1)}-\omega _{\beta
\alpha }|>\Gamma $ the contribution of the correlation term to the
variance of the matrix elements of $\rho_{\alpha \beta }$ can be
estimated in the ratio as
\begin{eqnarray}
R\equiv \frac{S_c}{S_d}\!\!\! &=&\!\!\!
-\frac{(n-2)\tilde K_1+\tilde K_2}{(n-1)K_1+K_2} \nonumber \\
&=&\!\!\! -\frac{(n-2)(m-n-1)(m-n+2)}{n(m-n)(m-n+3)}~.\label{final}
\end{eqnarray}
This equation shows that for $n=2$ we have $S_c=0$, which is easy to
check directly, since $\overline{H_{i'j'}H_{ij}}=0$
in this case. For $n>2$  the correlation contribution $S_c$ is
negative at the tails of the strength distribution. This means that
it indeed suppresses the transition amplitudes off-resonance (see
Fig.~1). For $n,m-n\gg 1$ the ratio $R$ is
approaching its limit value $-1$. It is easy to obtain from
Eq. (\ref{final}) that for $m-n\gg 1$
\begin{equation}\label{final1}
\frac{S_d+S_c}{S_d}=1+R\simeq \frac{2m}{n(m-n)}~.
\end{equation}
Thus, surprisingly, the role of the correlation contribution increases
with the number of particles.

For the numerical example shown in Fig. 1, $n=4$, $m=11$, one obtains:
$R=-0.39$, which means that the correlation contribution reduces
the magnitude of the squared matrix elements $\overline{M^2}$ between
compound
states almost by a factor of $2$ (for $\left| E^{(n_2)}-E^{(n_1)}-
\omega _{\beta \alpha }\right|>\Gamma $ ). The ratio found
numerically is $R\approx -0.45$ (Fig. 1b, $n_2=$150--250; larger
$n_2$ are probably too close to the boundary of the matrix for $R$ to
remain constant).

We would like to stress that the role of the correlation term does not
decrease with the increase of the numbers of particles and orbitals.
This prediction is supported by Fig. 2, which shows the behaviour of
the squared matrix element and its diagonal and correlation parts for
$n=7$ and $m=14$, $N=3432$. One can see that the suppression of the
matrix elements $M^2$ due to the correlation term at the tails is even
stronger than that in Fig. 1 [the numerically found ratio is
$R\approx -0.7$ vs $R=-0.55$ obtained from Eq. (\ref{final})]. The
correlation contribution should be even more important in compound
nuclei, where $N\sim 10^5$ . This case can be modeled by the
parameters $n=10$, $m=20$; then we have $R=-0.66$, or, equivalently,
$(S_d+S_c)/S_d=0.34$, which means that the correlations suppress the
squared element $M^2$ between compound states by a factor of $3$
(far from its maximum).

It is worth emphasizing that the existence of correlations due to the
perturbation theory admixtures of small components to the chaotic
eigenstates, which leads to a non-zero value of $S_c$ (\ref{SSc}), is
indeed non-trivial. For example, if one examines the summand of
Eq. (\ref{SSc}) as a function of $i$ and $j$, it would be hard to
guess that the sum itself is essentially non-zero, since positive and
negative values of $C_i^{(n_1)}C_j^{(n_1)} C_{i'}^{(n_2)}
C_{j'}^{(n_2)}$ seem to be equally frequent, and have roughly
the same magnitude, see Fig. 3.

Since $\sum_{n_1}S_c^{(n_1n_2)}=\sum_{n_2}S_c^{(n_1n_2)}=0$ (see
below), the suppression of $\overline{M^2}$ at the tails should be
accompanied by correlational enhancement of the matrix elements near
the maximum (at $\left|E^{(2)}-E^{(1)}-\omega _{\beta \alpha }
\right|<\Gamma $). Thus, we come to the important conclusion: even
for a random two-body interaction, the correlations produce some sort
of a ``correlation resonance'' in the distribution of the squared
matrix elements $M^2$. One should note that this increase of the
correlation effects in the matrix elements of a perturbation can be
explained by the increased correlations between the Hamiltonian matrix
elements when the number of particles and orbitals increases
($N/n\propto e^n)$.

Now we can estimate the size of the correlation contribution $S_c$
near the maximum of the the $M^2$ distribution (at $\left| E^{(n_2)}
-E^{(n_1)}-\omega _{\beta \alpha}\right| <\Gamma )$. First, we show
that after summation over one of the compound states, the correlation
contribution vanishes. Indeed,
$$
\sum_{n_2}S_c^{(n_1n_2)}=\sum_{n_2}\sum_{i\neq j,k\neq l}
C_i^{(n_1)}C_j^{(n_1)}C_k^{(n_2)}C_l^{(n_2)}\left\langle i\right| \rho
_{\alpha \beta }\left| k\right\rangle \left\langle l\right|
\rho _{\beta \alpha }\left| j\right\rangle
$$
\begin{equation}\label{zero}
=\sum_{i\neq j,k\neq l}C_i^{(n_1)}C_j^{(n_1)}\left\langle
i\right| \rho _{\alpha \beta }\left| k\right\rangle \left\langle l
\right|\rho _{\beta \alpha }\left| j\right\rangle
\sum_{n_2}C_k^{(n_2)}C_l^{(n_2)}=0~,
\end{equation}
where we take into account that the sum over $n_2$ in the expression
above is zero for $k\neq l$. Therefore, the negative value of
$S_c^{(n_1n_2)}$ at $\left|E^{(n_2)}-E^{(n_1)}-\omega _{\beta
\alpha }\right|>\Gamma $ must be compensated by its the positive
value near the maximum. The sum rule (\ref{zero}) allows one to make
a rough estimate of $S_c$ near the maximum of
$S_d$ (and $\overline{M^2}$).

Let us assume that $S_c=R_mS_d$ at
$|E^{(n_2)}-E^{(n_1)}-\omega _{\beta \alpha }|<\Gamma /2$, whereas
$S_c=R_tS_d$ at $|E^{(n_2)}-E^{(n_1)}-\omega _{\beta \alpha }|>
\Gamma /2$ [$R_t$ is given by Eq. (\ref{final})].
The distribution of $S_d^{(n_1n_2)}$ can be reasonably
approximated by the Breit-Wigner shape (see Secs. 1 and 4),
\begin{equation}\label{BW}
S_d^{(n_1n_2)}=\frac A{E^2+\Gamma ^2/4}~,
\end{equation}
where $E=E^{(n_2)}-E^{(n_1)}-\omega _{\beta \alpha }$, and $\Gamma =
\Gamma _{n_1}+\Gamma _{n_2}$. The sum rule (\ref{zero}) implies that
\begin{equation}\label{sumrule}
R_m\int\limits_0^{\Gamma /2}\frac{dE}{E^2+\Gamma ^2/4}
+R_t\int\limits_{\Gamma /2}^\infty \frac{dE}{E^2+\Gamma ^2/4}=0~.
\end{equation}
Since the two integrals in the above equation are equal, we have
$R_m=-R_t$.
Thus, near the maximum the correlation contribution $S_c$ is positive
and enhances the squared matrix element with respect to the diagonal
contribution,
\begin{equation}\label{estimate}
\frac{S_d+S_c}{S_d}=1+R_m=2-(1+R_t)\simeq 2\left[ 1-\frac {m}{n(m-n)}
\right] ~.
\end{equation}
Comparing the values of the ratio $S_c/S_d$ at the maximum and at the
tail in Fig. 1b ($n=4, m=11$), one can see that indeed,
$R_m\approx -R_t$. For larger $n$ and $m$ the correlation enhancement
factor asymptotically reaches its maximal value of 2. The numerical
example in Fig. 2 ($n=7,~m=14$) shows the enhancement of
$\overline{M^2}$ with respect to $S_d$ at the maximum even greater in
size than that predicted by Eq. (\ref{estimate}). This is not too
surprising since in Eqs. (\ref{BW})--(\ref{estimate}) we estimated
the average value of $R_m$ over an interval $\Delta E\simeq
\Gamma $ around the maximum rather than the peak value at the maximum.

A similar estimate of $S_c$ near maximum can be obtained by the direct
calculation of the small component contribution to $S_c$ (\ref{SSc}).
On an assumption that there are no correlations between principal
components of compound states we can separate the contribution of
small components. For example, in the resonance situation,
$E^{(n_2)}-E^{(n_1)}\simeq \omega _{\beta \alpha }$, if the
components $S_j^{(n_1)}$ and $S_{j'}^{(n_2)}$ are small
($|E_j-E^{(n_1)}|>\Gamma $, and consequently, $|E_{j'}-E^{(n_2)}|>
\Gamma $), then they contain contributions proportional to the
principal components $C_i^{(n_1)}$ and $C_{i'}^{(n_2)}$ [see Eqs.
(\ref{C1}), (\ref{C2})].
Analogously, $S_j^{(n_1)}$ and $S_{j'}^{(n_2)}$ may be among the
principal components, and then the small components $C_i^{(n_1)}$ and
$C_{i'}^{(n_2)}$ will contain correlated contributions. Thus, we have
the following estimate,
\begin{equation}\label{approx}
S_c^{(n_1n_2)}\simeq 2\widetilde{{\sum_{i}}'}{\sum _{{\rm small}\,j}}'
\left| C_i^{(n_1)}\right| ^2\left| C_{i'}^{(n_2)}\right| ^2
\frac{\overline{H_{ij}H_{i'j'}}}{(E^{(n_1)}-E_j)(E^{(n_2)}-E_{j'})}~.
\end{equation}
Since $E_{j'}-E_j=E_{i'}-E_i\simeq E^{(n_2)}-E^{(n_1)}$ for the
principal components $i$ and $i'$, $(E^{(n_1)}-E_j)$ and
$(E^{(n_2)}-E_{j'})$ in the denominator have always the same sign,
and $S_c$ is positive (recall that $\overline{H_{ij}H_{i'j'}}>0$).
Expression (\ref{approx}) can be estimated using the well known
formula for the spreading width, $\Gamma =2\pi \overline{H_{ij}^2}/D$,
where $D$ is the mean level spacing for the many-body states. This
yields $S_c\sim S_d$, in agreement with the previous estimate
(\ref{estimate}).

\section{Correlations between transition\protect \newline amplitudes}

We have shown that correlations between eigenvector components in a
system with a two-body interaction between particles must be taken
into account when calculating the variance of a matrix element between
compound states $\overline{M^2}$. Another question is whether the
above correlations between eigenstate components lead to correlations
between different matrix elements,
\begin{eqnarray}
M_{n_1n_2}M_{n_2n_3}&=&\left\langle n_1\right| \rho _{\alpha \beta }
\left|
n_2\right\rangle \left\langle n_2\right| \rho _{\beta \alpha }\left|
n_3\right\rangle \nonumber \\
&=&\sum C_i^{(n_1)}C_j^{(n_2)}C_k^{(n_2)}C_l^{(n_3)}\left\langle
i\right| \rho _{\alpha \beta }\left| j\right\rangle \left\langle k
\right| \rho _{\beta \alpha }\left| l\right\rangle ~.\label{MM}
\end{eqnarray}
Our analysis shows that the correlations of the type (\ref{MM}) are
absent, i.e., $\overline{M_{n_1n_2}M_{n_2n_3}}=\overline{M_{n_1n_2}}~
\overline{M_{n_2n_3}}=0$). The result could be different if the
principal components of different eigenvectors ($|n_1\rangle $ and
$|n_3\rangle $) were correlated. This effect takes place in the
separable interaction model \cite{F95,FG95}, but we have not found
such correlations in the TBRIM.

The absence of correlations between different amplitudes is confirmed
by direct numerical experiments. First, we have studied the
probability density of the
matrix elements $M_{n_1n_2}$ for different $n_1,n_2$ obtained for a
number of realizations of the two-body matrix elements
$V_{\alpha \beta \gamma \delta}$. Since the variance of
$M_{n_1n_2}$ depends on $n_1$ and $n_2$, the probability density of
$M_{n_1n_2}$ has been obtained by normalizing each matrix element
$M_{n_1n_2}$ to its root-mean-squared value which was calculated by
averaging over the realizations of $V_{\alpha \beta \gamma
\delta}$. The resulting probability density ${\cal P}\left(
M_{n_1n_2}/\sqrt{\overline{M_{n_1n_2}^2}}\right)$, averaged over
$N_r=5$ realizations of $V_{\alpha \beta \gamma\delta}$, turns out to
be quite close to Gaussian (Fig. 4).
This result follows from the fact that each matrix element between
compound states is the sum of a large number of random (or, almost
random) terms, see Eq. (\ref{n1n2}), so that the Central Limit Theorem
applies. Therefore, the correlations found in the previous section,
do not show up, unless more complicated correlations involving
{\em different} components of the {\em same} eigenstate, like those
in Eqs. (\ref{Hcorr}) or (\ref{M2}), are probed.

To check whether some correlations between different matrix elements
(\ref{MM}) exist, we have plotted the matrix element
$\langle n_1|\rho _{\alpha \beta }| n_2\rangle $ versus another one,
$\langle n_1| \rho _{\alpha \beta }|n_2-1\rangle $, where
$|n_2-1\rangle $ is the eigenstate immediately preceding
$|n_2\rangle $, for some fixed $n_1$, $n_2$, $\alpha $, and
$\beta $, obtained from $N_r=387$ different Hamiltonian matrices
(Fig. 5a). The latter were generated by using different random
realizations of $V_{\alpha \beta \gamma\delta}$. Detailed analysis of
the distribution of the points in this figure does not reveal any
sort of correlations.

The next question is the existence of correlations between matrix
elements $M_{n_1n_2}$ and $W_{n_1n_2}$ of different operators for the
same compound states $|n_1\rangle $ and $|n_2\rangle $. If the
expansions of these matrix elements [see Eq. (\ref{n1n2})] contain
identical matrix elements of the density matrix operator
$\rho _{\alpha \beta }$, such correlations, in principal, do exist:
\begin{equation}\label{MW}
\overline{M_{n_1n_2}W_{n_2n_1}}=\sum _{\alpha \beta }M_{\alpha \beta }
W_{\beta \alpha }\overline{\langle n_1|\rho _{\alpha \beta }|n_2
\rangle ^2} \neq 0~.
\end{equation}
A more complicated question is whether the matrix elements of
different ``elementary'' transition operators $\rho _{\alpha \beta }$
and $\rho _{\gamma \delta }$ are indeed uncorrelated [as we assumed
writing Eq. (\ref{MW})]. The product of such two matrix elements can
be presented in the form
\begin{eqnarray}\label{corrMW}
\langle n_1| \rho _{\alpha \beta }| n_2\rangle \langle n_2|
\rho _{\delta \gamma }| n_1\rangle &\!\!=&\!\!
\sum_{ijkl}C_i^{(n_1)}C_j^{(n_1)}C_k^{(n_2)}C_l^{(n_2)}
\langle i| \rho _{\alpha \beta }| k\rangle \langle l|\rho _{\delta
\gamma } |j\rangle \\
&\!\!\simeq &\!\!
2\widetilde{{\sum_{ij''}}'}\frac{\left| C_i^{(n_1)}\right|^2\left|
C_{j''}^{(n_2)}\right| ^2
H_{ij}H_{i'j''}}{(E^{(n_1)}-E_j)(E^{(n_2)}-E_{i'})}~,\label{corrMW1}
\end{eqnarray}
where the last expression is written for $n_1$ and $n_2$ far from the
maximum ($|E^{(n_2)}-E^{(n_1)}-\omega _{\beta \alpha }|>\Gamma ,~
|E^{(n_2)}-E^{(n_1)}-\omega _{\delta \gamma }|>\Gamma $), and
$|i'\rangle = a_\beta ^\dagger a_\alpha |i\rangle $, $|j''\rangle =
a_\delta ^\dagger a_\gamma |j\rangle $. It can be shown that in our
model $\overline{H_{ij}H_{i'j''}}\propto \delta _{\alpha \gamma }
\delta _{\beta \delta }$. Therefore, there are no terms in the
expression (\ref{corrMW1}) which would give non-zero contributions,
and the average of (\ref{corrMW}) is zero. The absence of
correlations in this case is illustrated by Fig. 5b, where numerical
data obtained in the TBRIM are presented. As in the case of the matrix
elements between different pairs of compound states, no correlations
can be seen between the matrix elements of different transition
operators.

\section{Statistical description of the transition amplitudes}

In this Section we use the TBRIM to test the validity of the
statistical approach to the calculation of transition amplitudes
between compound states of complex systems developed in
\cite{F94,FGGK94,FV93}. In what follows we first outline
the main ideas of the statistical approach. The variance of the
matrix elements of an operator $\hat M$ (\ref{M}) between the compound
states $|n_1\rangle $ and $|n_2\rangle $ can be presented in the
following form [compare with Eq. (\ref{MW})],
\begin{equation}\label{Mrho}
\overline{\left| M_{n_1n_2}\right| ^2}=\sum_{\alpha \beta}
\left| M_{\alpha \beta }\right| ^2\overline{\left|\langle n_1|
\rho _{\alpha \beta}|n_2\rangle \right|^2}~,
\end{equation}
where we have taken into account the result of the previous section
that the average of the correlator (\ref{corrMW}) is zero unless
$\gamma =\alpha ,~\delta =\beta $. Therefore, the calculation of
$\overline{\left| M_{n_1n_2} \right| ^2}$ (or $\overline{M_{n_1n_2}
W_{n_2n_1}}$) is reduced to the calculation of
$\overline{|\langle n_1|\rho _{\alpha \beta}|n_2\rangle |^2}$.

It was suggested in Sec. 2 that $\overline{|\langle n_1|\rho _{\alpha
\beta}|n_2\rangle |^2}$ can be presented as the sum of the diagonal
and correlational sums $S_d$ and $S_c$, Eqs. (\ref{M2})--(\ref{Sc}).
Since we have already estimated the ratio $(S_d+S_c)/S_d$, it is
enough to calculate only $S_d$, Eq. (\ref{Sd}).
Following \cite{FGGK94} let us replace the squared components
$\left| C_i^{(n_1)}\right| ^2$ and $\left| C_k^{(n_2)}\right| ^2$ by
their average values,
\begin{equation}\label{w}
\overline{\left| C_i^{(n_1)}\right| ^2}\equiv w(E_i,E^{(n_1)})~,\quad
\overline{\left| C_k^{(n_2)}\right| ^2}\equiv w(E_k,E^{(n_2)})~,
\end{equation}
where the averaging goes as usual either over a number of realizations
of the two-body interaction matrix elements (ensemble average), or
over a number of neighbouring eigenstates (physical energy average);
in the spirit of ergodicity the results are presumably the same. The
function $w$ is proportional to the strength function introduced by
Wigner \cite{wigner}, which is also called the local spectral density
of states. Note that definition (\ref{w}) also implies that the
mean-square contribution of the component $i$ in the eigenstate $n_1$
is determined by their energies, $E_i$ and $E^{(n_1)}$ (in fact, by
their difference $|E_i-E^{(n_1)}|$). For states localized  in the
given basis, $w$ is a bell-shaped function with a typical width
determined by the spreading width $\Gamma $. There is some theoretical
and experimental evidence that it can be approximated by the
Breit-Wigner formula, although its tails decrease faster then
$|E_i-E^{(n_1)}|^{-2}$ (see references in Sec. 1).

The diagonal sum now takes the form
\begin{equation}\label{Sdw}
S_d^{(n_1n_2)}=\sum_{ik}w(E_i,E^{(n_1)})w(E_k,E^{(n_2)})\langle i|
\rho _{\alpha \beta }|k\rangle \langle k|\rho _{\beta\alpha }|i
\rangle ~.
\end{equation}
The summation over $k$ for a fixed $i$ includes only one state,
$|k\rangle = \rho _{\beta\alpha }|i\rangle $, with $E_k=E_i+
\omega _{\beta \alpha }$. On the other hand, we can write
\begin{equation}\label{nn}
\sum_k\langle i|\rho _{\alpha \beta }|k\rangle \langle k|\rho _{\beta
\alpha }|i\rangle =\langle i|\rho _{\alpha \beta }\rho _{\beta
\alpha}|i\rangle =\langle i|\hat n_\alpha (1-\hat n_\beta )| i
\rangle ~,
\end{equation}
where $\hat n_\alpha =a_\alpha ^\dagger a_\alpha $ and $\hat n_\beta
=a_\beta ^\dagger a_\beta $ are the occupation number operators. Thus,
we obtain,
\begin{equation}\label{rhofinal}
S_d^{(n_1n_2)}=\sum_iw(E_i,E^{(n_1)})w( E_i+\omega _{\beta \alpha },
E^{(n_2)})\langle i|\hat n_\alpha (1-\hat n_\beta )|i\rangle ~.
\end{equation}

The matrix element $\langle i|\hat n_\alpha (1-\hat n_\beta)| i
\rangle $ is equal to 1 if the orbital $\alpha $ is occupied and
$\beta $ is vacant in the basis state $|i\rangle $, otherwise, it is
zero. We used this fact earlier [Eq. (\ref{SSd})] to reduce the
summation to these states only. Now we proceed in a different way.
Both $w$'s in (\ref{rhofinal}) are smooth functions of energy
normalized as $\sum _iw(E_i,E^{(n_1)})=1$. This allows one to replace
the matrix element of $\hat n_\alpha (1-\hat n_\beta )$ by its
expectation value,
\begin{eqnarray}
\overline{\langle i|\hat n_\alpha (1-\hat n_\beta)| i\rangle }=
\sum _i w(E_i,E^{(n_1)})\langle i|\hat n_\alpha (1-\hat n_\beta)|
i\rangle \nonumber \\
=\sum _i\overline {\left| C_i^{(n_1)}\right| ^2}\langle i|
\hat n_\alpha (1-\hat n_\beta)| i\rangle \simeq \langle \hat n_\alpha
(1-\hat n_\beta)\rangle _{n_1}~.\label{mean}
\end{eqnarray}
The sign $\simeq $ above is a reminder that the left-hand side is the
local average over the states $|n_1\rangle $. Practically, when the
number of components is large, the fluctuations of
$\langle \hat n_\alpha (1-\hat n_\beta )\rangle _{n_1}$ are expected
to be small. Now we can re-write Eq. (\ref{rhofinal}) in a form
similar to Eq. (\ref{SSd}), but without any restrictions on the
summation variable $i$,
\begin{equation}\label{end}
S_d^{(n_1n_2)}=\langle \hat n_\alpha (1-\hat n_\beta)\rangle _{n_1}
\sum_iw(E_i,E^{(n_1)})w( E_i+\omega _{\beta \alpha },E^{(n_2)})~.
\end{equation}

It was shown in \cite{FGGK94} that under some reasonable assumptions
about the functions $w$ one can introduce a ``spread
$\delta $-function'' $\tilde \delta (\Delta )$,
\begin{eqnarray}
\tilde \delta (\Delta )&\!\!\!=&\!\!\! D_2^{-1}\sum_iw(E_i,E^{(n_1)})
w( E_i+\omega _{\beta \alpha },E^{(n_2)})\nonumber \\
&\!\!\!= &\!\!\! D_2^{-1}\int \frac{dE_i}{D_1}
w(E_i,E^{(n_1)})w( E_i+\omega _{\beta \alpha },E^{(n_2)})~,\label{del}
\end{eqnarray}
where $\Delta =E^{(n_2)}-E^{(n_1)}-\omega _{\beta \alpha }$, $D_1$ and
$D_2$ are local mean level spacings for the $n_1$ and $n_2$
eigenstates. The function $\tilde \delta (\Delta )$ is symmetric, its
characteristic width is determined by the spreading widths of the
eigenstates $n_1$ and $n_2$, $\Gamma \sim \Gamma _1+\Gamma _2$, and
it is normalized to unity, $\int \tilde \delta (\Delta )d\Delta =1$,
just as the standard $\delta $-function. If $w$'s have Breit-Wigner
shapes, $\tilde \delta $ is also a Breit-Wigner function with
$\Gamma =\Gamma _1 +\Gamma _2$. The fact that $S_d^{(n_1n_2)}$ is
proportional to the function $\tilde \delta (\Delta )$ is a
particular manifestation of the energy conservation for transitions
between the quasistationary basis states \cite{FV93} [if $\Gamma
\rightarrow 0$, then $\tilde \delta (\Delta ) \rightarrow
\delta (\Delta )$]. Using Eqs. (\ref{Mrho}), (\ref{end}), and
(\ref{del}) we can finally present the diagonal contribution to the
variance of the matrix element $M_{n_1n_2}$
in the following form
\begin{equation}\label{diaM}
\overline{\left| M_{n_1n_2}\right| ^2}_{\rm diag}=\sum_{\alpha \beta}
\left| M_{\alpha \beta }\right| ^2 \langle \hat n_\alpha
(1-\hat n_\beta) \rangle _{n_1}D_2\tilde \delta (E^{(n_2)}-E^{(n_1)}
-\omega _{\beta \alpha })~.
\end{equation}
This expression is apparently asymmetric with respect to the states
$n_1$ and
$n_2$. By performing the calculation in a different way we can obtain
\begin{equation}\label{end1}
S_d^{(n_1n_2)}=\langle \hat n_\beta (1-\hat n_\alpha)\rangle _{n_2}
\sum_kw(E_k-\omega _{\beta \alpha },E^{(n_1)})w(E_k,E^{(n_2)})~,
\end{equation}
instead of Eq. (\ref{end}), and thereby arrive at a different formula
for the variance,
\begin{equation}\label{diaM1}
\overline{\left| M_{n_1n_2}\right| ^2}_{\rm diag}=\sum_{\alpha \beta}
\left| M_{\alpha \beta }\right| ^2 \langle \hat n_\beta
(1-\hat n_\alpha ) \rangle _{n_2}D_1\tilde \delta (E^{(n_2)}-E^{(n_1)}
-\omega _{\beta \alpha })~,
\end{equation}
where the occupancies factor is now calculated for the state $n_2$ (it
represents the probability to find the orbital $\beta $ occupied, and
$\alpha $ empty). If the suppositions made in the above derivations
are correct, the two formulae (\ref{diaM}) and (\ref{diaM1}) should
give identical results.

In the present work we use the TBRIM to check the accuracy of the
statistical approach described above. Figure Fig. 6a presents a
comparison between the values of $S_d^{(n_1n_2)}$ as given by Eqs.
(\ref{end}), (\ref{end1}), and those from the initial expression
(\ref{Sd}). Clearly, there is a good agreement between the three
formulae.

It is quite important for applications of the statistical approach
(see \cite{FG95,FV93}) that further simplifications be made by
replacing the correlated occupancies product $\langle \hat n_\alpha
\hat n_\beta \rangle _{n_1}$ in Eq. (\ref{mean}) by the product of the
two mean values, $\langle \hat n_\alpha \rangle _{n_1}\langle
\hat n_\beta \rangle _{n_1}$. This is definitely a valid operation
when the numbers of excited particles and active orbitals are large,
so that the occupation numbers for different orbitals become
statistically independent. Then one would be able to use the
following relation
\begin{equation}\label{occup}
\left\langle \hat n_\alpha (1-\hat n_\beta )\right\rangle \simeq 
n(\epsilon _\alpha )[1-n(\epsilon _\beta )]~, 
\end{equation}
where $n(\epsilon _\alpha )$ and $n(\epsilon _\alpha )$ are the
occupation numbers. (They can be calculated, e.g., using the
Fermi-Dirac formula with an effective temperature, see
\cite{FIC95,FV93}; see also \cite{zelev} where the relation between
thermalization and chaos is studied in nuclear shell-model
calculations). The result of such simplification is shown in Fig. 6b,
where the diagonal contribution (\ref{Sd}) is again compared with the
values obtained from Eqs. (\ref{end}), (\ref{end1}), using
approximation (\ref{occup}). In spite of the fact that the TBRIM
calculation included $n=4$ particles only, the agreement remains
quite reasonable, the error being about 10\%. To examine the quality
of the approximation at the tails of the distribution, Fig. 7 shows
the ratio of $S_d$ as given by Eqs. (\ref{end}), (\ref{end1}) to the
directly calculated diagonal term $S_d$, Eq. (\ref{Sd}). The
difference between Fig. 7a and Fig. 7b highlights the inaccuracy
introduced by an additional approximation (\ref{occup}) for the
occupation numbers.

In order to make a more direct test of the validity of substitution
(\ref{occup}), we plotted in Fig. 8 the correlator
$\langle \hat n_\alpha \hat n_\beta \rangle _{n_1}/[\langle
\hat n_\alpha \rangle _{n_1}\langle \hat n_\beta \rangle _{n_1}]$ as
function of $n_1$. Consistent with the small number of particles,
this correlator displays large fluctuations, however, its average
value of about 0.8 is still rather close to 1.

\section{Spreading widths for different basis components}

In Section 4 when considering the statistical approach to the
calculation of the variance of matrix elements between compound
states, it was assumed that the spreading widths $\Gamma $ are the
same for all basis components. However, this question is not trivial.
As is discussed in the literature, the spreading widths of components
corresponding to different number of excited particles could have
significantly different values. For example, in \cite{hussein} it is
argued that two-particle-one-hole states ($2p$-$1h$) can lead to
correlations between values of parity nonconserving effects
\cite{FR91},
if the spreading width of the $2p$-$1h$ states is two orders of
magnitudes smaller than that of $1p$ states. In such case one might
expect that in our model the spreading width showed a rapid decrease
as function of the number of excited particles in the basis component.

To study this question in detail, we have performed additional tests.
In Fig. 9 the root-mean-squared spreading width $\Gamma _j$ for all
basis states $| j\rangle $ is presented for $n=6$, $m=12$ ($N=924$).
Here we use the following definition:
\begin{equation}\label{gamma}
\Gamma _j^2=\langle j|\left( H-\langle j|H|j\rangle \right) ^2|
j\rangle ~, \end{equation}
which can also be presented in several equivalent forms,
\begin{eqnarray}\label{gamma1}
\Gamma _j^2\!\!\! &=&\!\!\!\left( H^2\right) _{jj}-\left( H_{jj}
\right) ^2 \\
\!\!\!&=&\!\!\!\sum _{i\neq j}H_{ij}^2 \label{gamma2} \\
\!\!\!&=&\!\!\! M_2-\left( M_1\right) ^2 ~,\label{gamma3}
\end{eqnarray}
the last one relating $\Gamma _j$ directly to the moments $M_p$ of the
strength function
$\rho _w(E,j)\equiv \sum _n\left| C_j^{(n)}\right| ^2\delta
(E- E^{(n)})$,
\begin{equation}\label{mom}
M_p=\int \rho _w(E,j)\,E^pdE=\sum _n\left| C_j^{(n)}\right| ^2
\left( E^{(n)}\right) ^p~.
\end{equation}
Equations (\ref{gamma1})--(\ref{gamma3}) can be obtained using
closure, $\sum _j|j\rangle \langle j|=\sum _n|n\rangle \langle n|=1$,
where $|j\rangle$ and $|n\rangle $ are the basis state and the
eigenstate respectively, $\langle j|n\rangle \equiv C_j^{(n)}$.
We should note that the r.m.s. spreading width is different from that
introduced intuitively in Sec. 4 as the characteristic width of the
strength function. For example, if the strength function has a
Breit-Wigner form, its second and higher moments are infinite. In the
Wigner's BRM the r.m.s. spreading width is determined by the
bandwidth $b$ as $\Gamma _j=\sqrt{2bV^2}$, whereas the ``Breit-Wigner
spreading width'' is $\Gamma _{\rm BW}= 2\pi v^2/D$ (see
Introduction). However, in a more realistic situation the
strength function drops rapidly, its second moment is finite, and the
difference between the r.m.s. $\Gamma $ and $\Gamma _{\rm BW}$ is not
large.

From Figure 9 one can see that apart from small natural fluctuations,
the r.m.s. spreading width is the same for all components. To exclude
a weak dependence of $\Gamma _j$ on the energy of the basis state $j$
(boundary effects seen as rises of $\Gamma _j$ at small and large
$j$), we calculate the mean value $\overline{\Gamma }$ and the r.m.s
deviation $\delta \Gamma $ of the spreading width by averaging over
$j=$50--874. The results are: $\overline{\Gamma }\approx 2.22$,
$\delta \Gamma \approx 0.12$. The latter value shows that the
fluctuations of the width are very small. This result is
in agreement with computations made for the Ce atom \cite{FGGK94};
similar results have been recently obtained in the nuclear $sd$-shell
model calculations \cite{zelev,report}. The fluctuations
of the width are small due to the large number of ``decay channels''
for each basis component (each component is coupled to many others by
random interaction) \cite{FG95}. Formally, this can be obtained from
Eq. (\ref{gamma2}). For example, in the TBRIM one obtains
\begin{equation}\label{gamran}
\overline{\Gamma ^2}=(n-1)V^2K_1+V^2K_2=K_{12}V^2~,
\end{equation}
which shows that for large number of ``independent decay channels'',
$K_{12}= (n-1)K_1+K_2$ [see Eq. (\ref{K123})] the statistics of
$\Gamma _j$ is given by the $\chi ^2$ distribution with $K_{12}\gg 1$
degrees of freedom, resulting in the $\sim 1/\sqrt{K_{12}}$ decrease
of fluctuations. More accurately, the relative r.m.s. fluctuation of
the squared width (\ref{gamma}) is given by $\delta (\Gamma ^2)/
\sqrt{\overline{\Gamma ^2}}=\sqrt{2/K_{12}}$. For $n=6$, $m=12$ Eq.
(\ref{gamran}) yields $\sqrt{\overline{\Gamma ^2}} \approx 2.41$, and
the relative fluctuation is 0.07. These values are close to
the numerical ones quoted above (the discrepancy is mainly due to the
difference between the mean width and its r.m.s. value, and the
corresponding difference in the fluctuations of $\Gamma _j$ and
$\Gamma _j^2$).

To check the independence of $\Gamma _j$ of the number of excited
particles in the component $j$, we have calculated the average
spreading width $\overline {\Gamma (p)}$ for basis states with a
fixed number of excited particles $p$, $p=1,2,...,n-1$. In the
numerical experiment shown in Fig. 6 all $\overline {\Gamma (p)}$ for
$p=$1--5 proved to be approximately the same,
$\overline {\Gamma (p)}\approx 2.2$.

The above consideration shows that the statistical approach does not
provide any support for the dependence of the spreading width on the
number of excited particles. This indicates that the argumentation in
favour of a strong dependence, based on different decay phase volumes
for different numbers of excited particles, seems to be incorrect. In
our opinion, the difference in the spreading widths could appear as a
result of specific dynamical effects. For example, this could be an
influence of levels in other potential wells which appear at higher
nuclear deformation, or due to an interaction with collective motions,
such as rotations and vibrations.

\section{Conclusions}
The calculation of the mean square matrix element of an operator
between compound states of a many-body system has been considered.
We have shown that the two-body nature of the interaction between
particles manifests itself in the existence of correlations between
the components of the ``chaotic'' compound eigenstates. These
correlations taken together with the correlations between the
many-particle Hamiltonian matrix elements result in a relatively large
correlation contribution to the mean square matrix element. The
correlations exist even if the two-body matrix elements are
independent random variables, as in the TBRIM.
Such correlations can be understood in terms of the perturbative
mixing of the distant small components to the principal components of
the eigenstates. If the Hamiltonian matrix elements are random
variables the correlations of this type vanish.

One of the most interesting feature of the correlations found in our
work is that they do not decrease with the increase of the number of
excited particles or active orbitals. Thus, they must be taken into
account when calculating matrix elements of a weak interaction between
compound states in nuclei.

Another feature concerns the shape of the distribution of the density
matrix operator near its maximum. As one can see from Figs. 1a,2a, the
correlations create a sharp spikelike form of the distribution,
instead of a smooth Gaussian or Breit-Wigner form. With such sharp
peaks, the strength function for any particular operator $\hat M$ can
have the so-called gross-structure, due to many single particle
transition terms in the expression (\ref{Mrho}). Without these
specific correlations, the strength function would be much smoother
and the gross-structure would not be seen. It is also
interesting to note that there are very large mesoscopic-type
fluctuations in the distribution near the maximum, depending on a
specific (random) realization of the two-body interaction $V$ . This
fact is also the consequence of strong correlations.

Our study also demonstrated that the spreading widths of different
basis components are approximately constant and fluctuate very weakly.
In particular, we have not found any dependence on the number of
excited particles in the component.

The statistical approach to the calculations of such matrix elements
has been tested in the present work with the help of the TBRIM. The
numerical results obtained in this work support the validity of the
statistical approach. The TBRIM has also enabled us to check that the
matrix elements of different transition operators between a pair of
compound states are uncorrelated, as are the matrix elements of a
given operator between different pairs of compound states.

\renewcommand{\thesection}{}
\section{Acknowledgments}

The authors are grateful to O. P. Sushkov and M. G. Kozlov for
valuable discussions. This work was supported by the Australian
Research Council and Gordon Godfrey Fund.

\newpage


\newpage

\begin{center}
FIGURE CAPTIONS
\end{center}

Figure 1.\\

(a) Mean square matrix element (\ref{M2}) calculated in the TBRIM for
$n=4$ particles and $m=11$ orbitals, $\alpha =4$, $\beta =5$, as a
function of the eigenstate $n_2$ for $n_1=55$. Averaging over
$N_r=100$ Hamiltonian matrices $H_{ij}$ for different realizations of
the random two-body matrix elements has been made. Dots correspond to
the sum $S_d+S_c$ while the solid line represents the diagonal
contribution $S_d$ only [see (\ref{Sd})].\\

(b) Ratio $R=S_c/S_d$ of the correlation contribution to the diagonal
contribution.\\

Figure 2.\\

Same as in Figure 1, for $n=7$, $m=14$, $\alpha =7$, $\beta =8$. The
data obtained for a single Hamiltonian matrix of the size $N=3432$;
$n_1=575$. Note the increased role of the correlation contribution
$S_c$.\\

Figure 3.\\
The distribution of the items of the sum (\ref{SSc}), $\xi_{ij}=
C_i^{(n_1)}C_j^{(n_1)}C_{i'}^{(n_2)}C_{j'}^{(n_2)}$, for $n_1=55$,
$n_2=66$, obtained in the TBRIM
for the same set of parameters as in Fig. 1, averaged over $N_r=100$
realizations of $V$. Indices $i$ and $j$ in the figure run over those
84 components in which $\alpha $ is occupied and $\beta $ is vacant.\\

a. Positive values.\\

b. Negative values (absolute values).\\
 
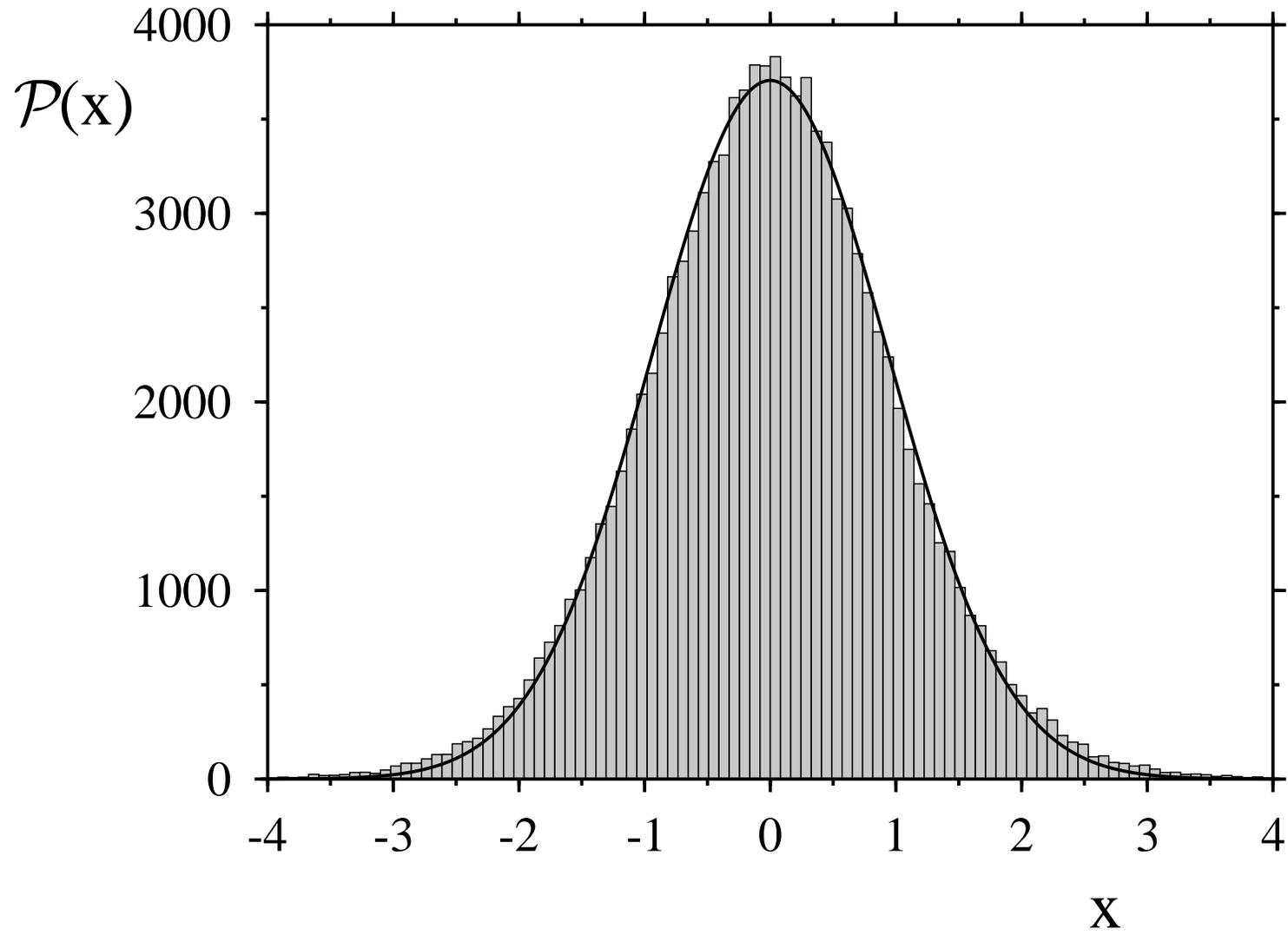
Figure 4.\\

Probability density of the normalized matrix elements
$x\equiv \langle n_1|\rho _{\alpha \beta }| n_2\rangle /\left(
\overline{\left|\langle n_1|\rho _{\alpha \beta }| n_2\rangle
\right | ^2} \right) ^{1/2}$ in the TBRIM for the parameters of
Fig. 1. The histogram is obtained for $N_r=5$ Hamiltonian matrices.
Solid curve is the normalized Gaussian distribution.\\

Figure 5.\\

(a) The matrix element $y_{\alpha \beta }=\left\langle n_1\right| \rho
_{\alpha \beta }\left| n_2\right\rangle $ for $n_1=55$, $n_2=67$,
plotted vs. the matrix element $x_{\alpha \beta }=\left\langle n_1
\right| \rho _{\alpha \beta }\left| n_3\right\rangle $ for $n_3=66$;
$\alpha =4$, $\beta =5$ and other TBRIM parameters as in Fig. 1. The
number of points in the figure is $N_r=387$. No evidence of
correlations between $x$ and $y$ is present.\\

(b) The matrix element $y_{\alpha \beta }=\left\langle n_1\right| \rho
_{\alpha \beta }\left| n_2\right\rangle $ for $\alpha =4$, $\beta =6$
vs. $x_{\alpha \gamma }=\left\langle n_1\right| \rho _{\alpha \gamma }
\left| n_2\right\rangle $ with $\gamma =5$ for $n_1=55$ and $n_2=66$.
Again, there is no indication of correlations between $x$ and $y$.
Note the difference in the vertical and horizontal scale due to the
fact that for given $n_1$ and $n_2$ the energy difference
$E^{(n_1)}-E^{(n_2)}$ is approximately in-resonance for the
transition between $\alpha =4$ and $\beta =5$ and off-resonance for
$\alpha =4$ and $\gamma =6$.\\

Figure 6.\\

(a) The diagonal contribution to the mean-square matrix element as
obtained from Eqs. (\ref{end}), (\ref{end1}) (solid lines) in
comparison with the direct calculation of $S_d$, Eq. (\ref{Sd})
(circles). The TBRIM parameters are the same as in Fig.~1.\\

(b) Same as (a), with the occupancy factors in Eqs. (\ref{end}),
(\ref{end1}) calculated by means of Eq. (\ref{occup}).\\

Figure 7.\\

(a) The ratio $R_a$ of the approximation represented in Fig. 6a by
the solid lines, to the value of $S_d$. \\

(b) Same as in Fig. 7b for the data of Fig. 6b.\\

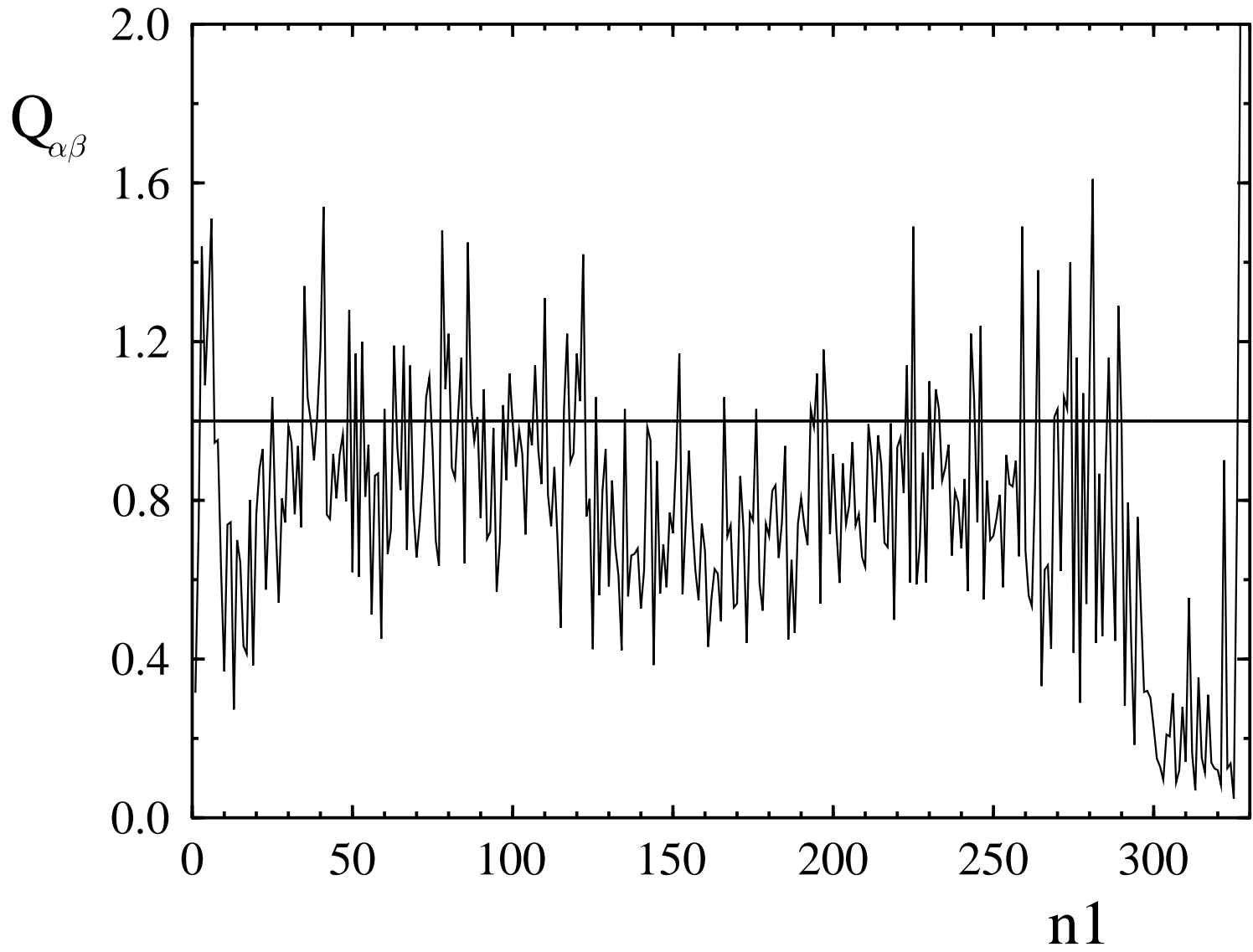
Figure 8.\\

The correlator $Q_{\alpha \beta }\equiv
\langle \hat n_\alpha \hat n_\beta \rangle _{n_1}/
[\langle \hat n_\alpha \rangle _{n_1}\langle\hat n_\beta
\rangle _{n_1}]$ vs the eigenstate number $n_1$. The TBRIM parameters
are the same as in Fig.~1. The average value of the correlator is
about 0.8, which means that the correlations between the occupancies
of different orbitals are not very strong.\\

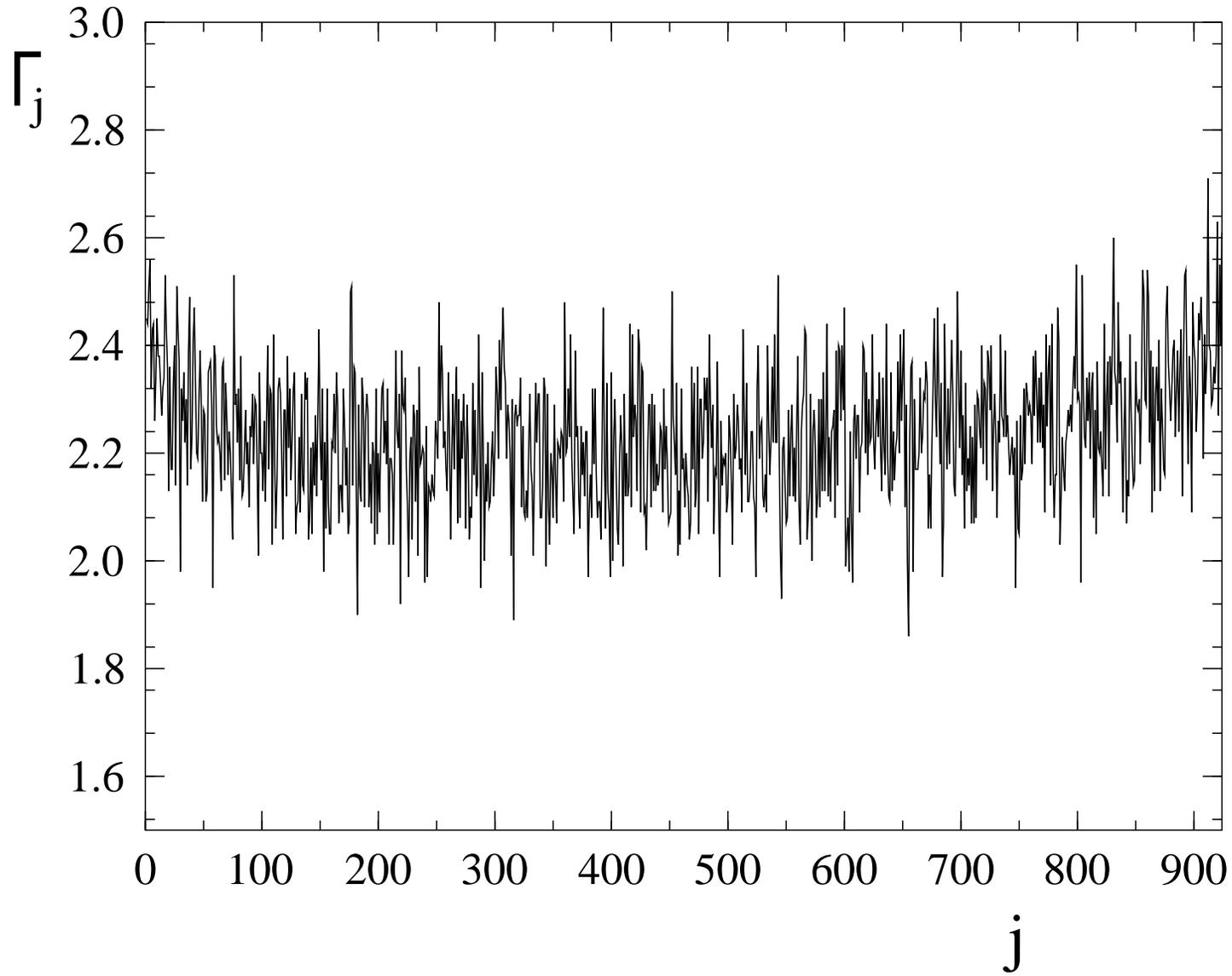
Figure 9.\\

The spreading width $\Gamma _j$ calculated as the r.m.s. deviation
from the center of the distribution of the components $\left|
C_j^{(n_1)}\right| ^2$ for each basis state $j$. The data are
obtained for one matrix $H_{ij}$ corresponding to $n=6$ particles and
$m=12$ orbitals.

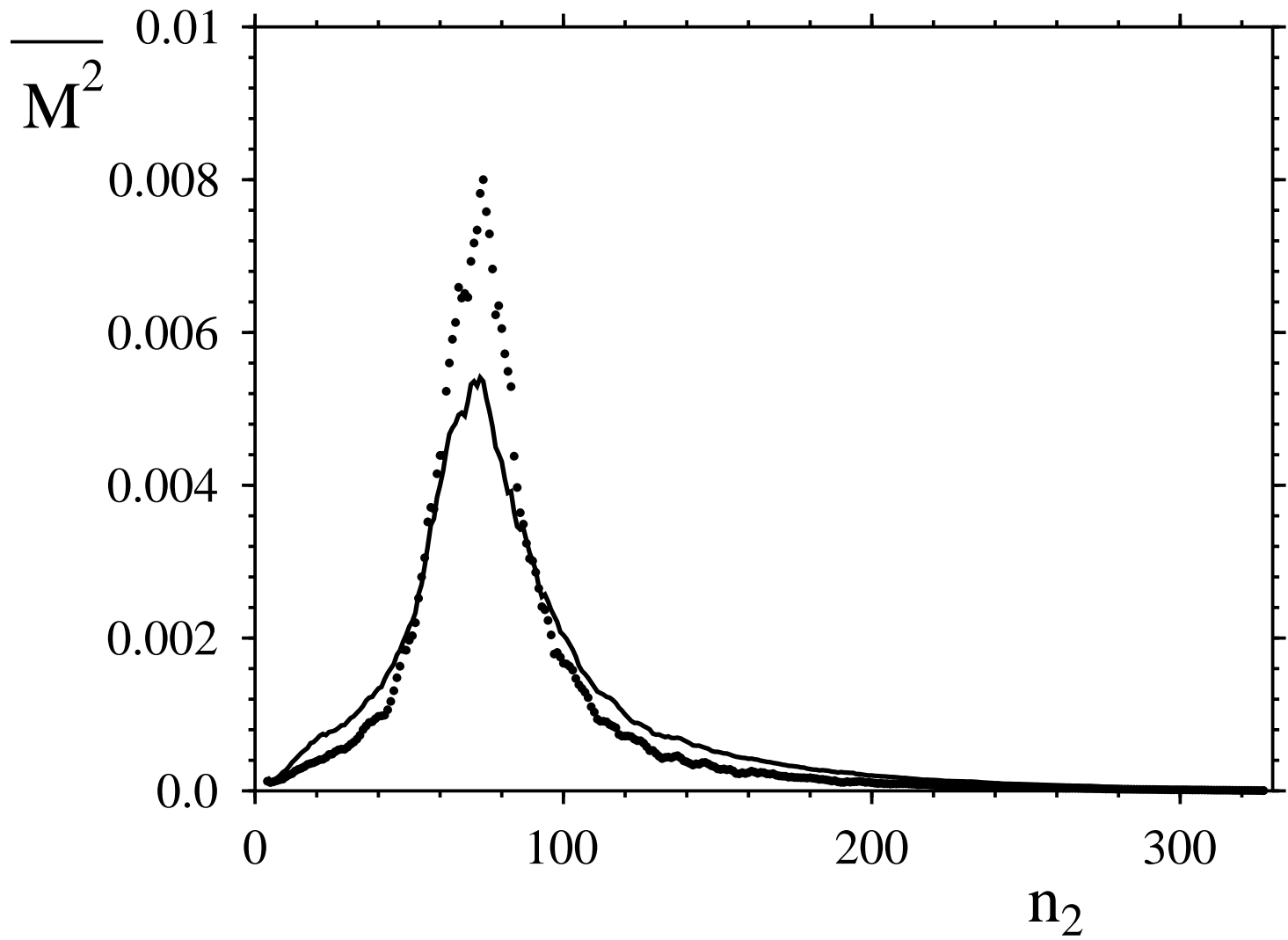
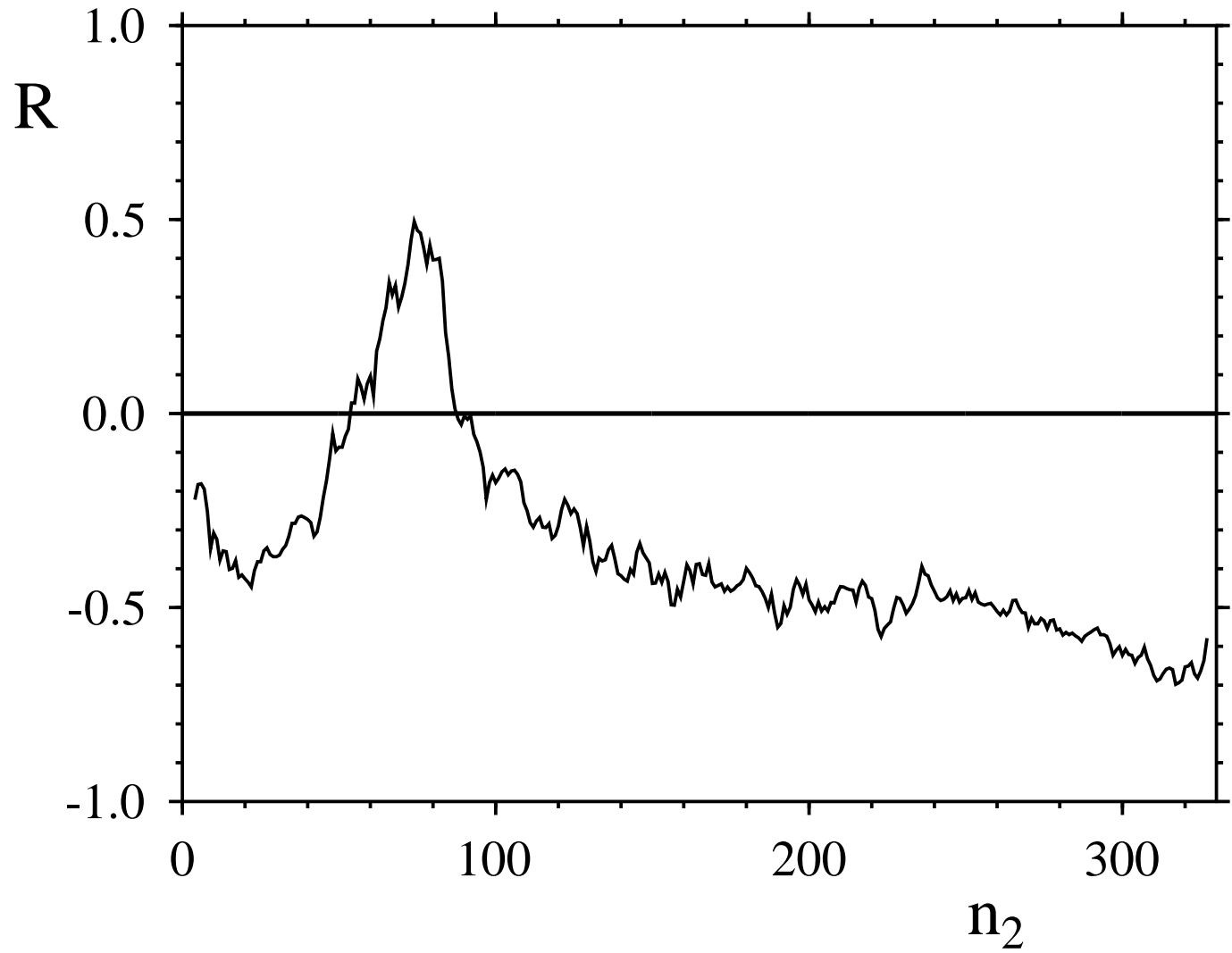
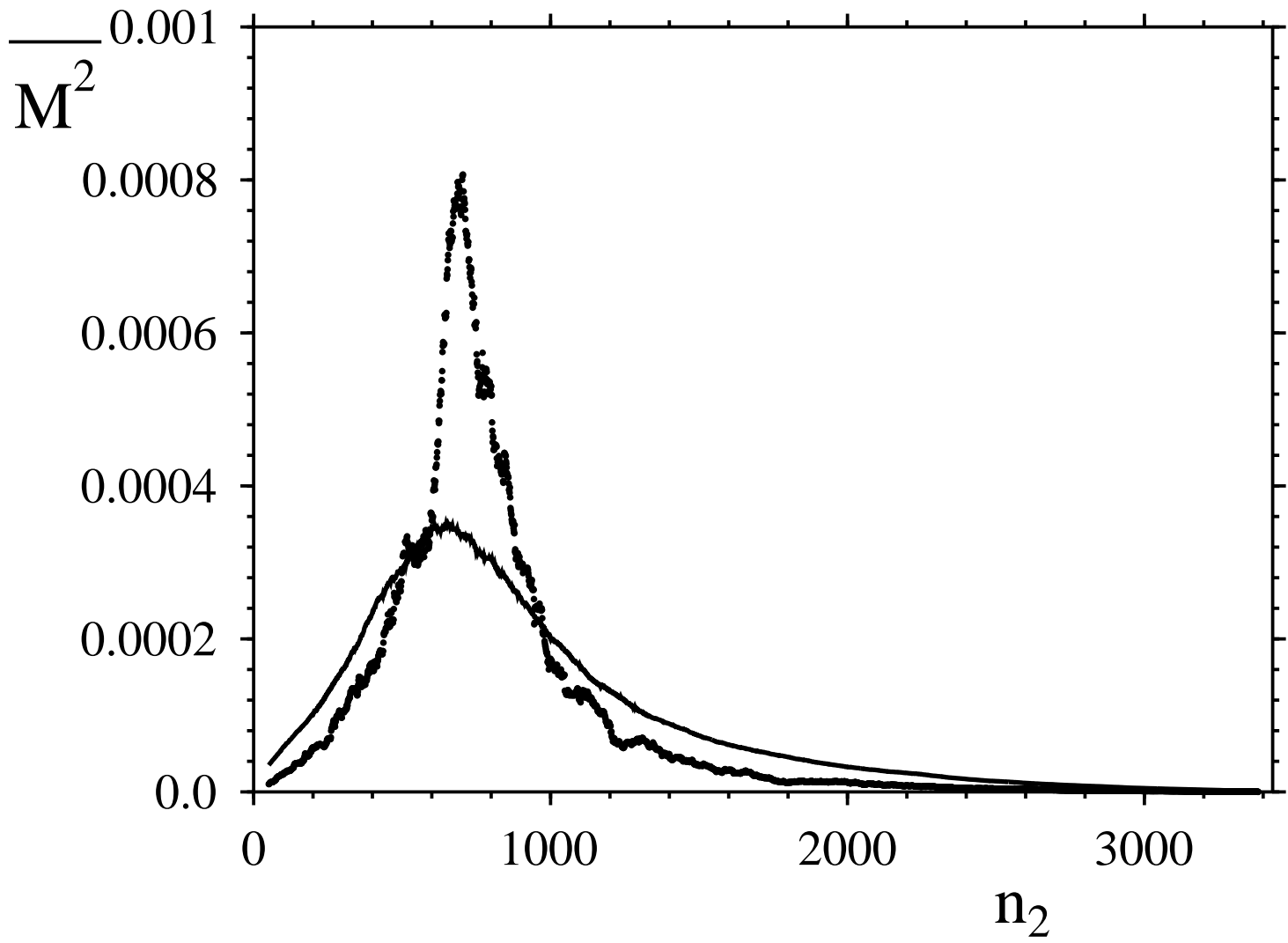
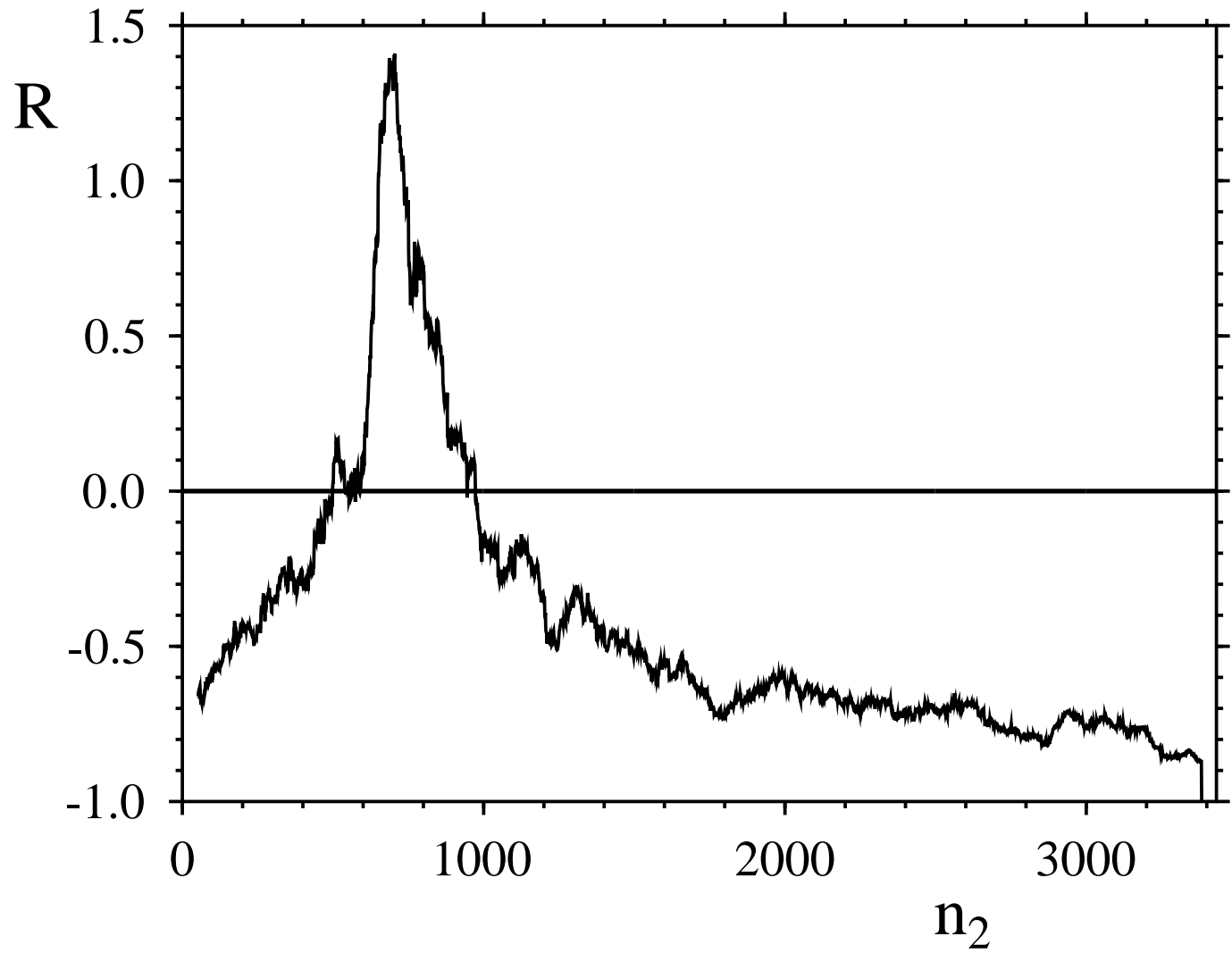
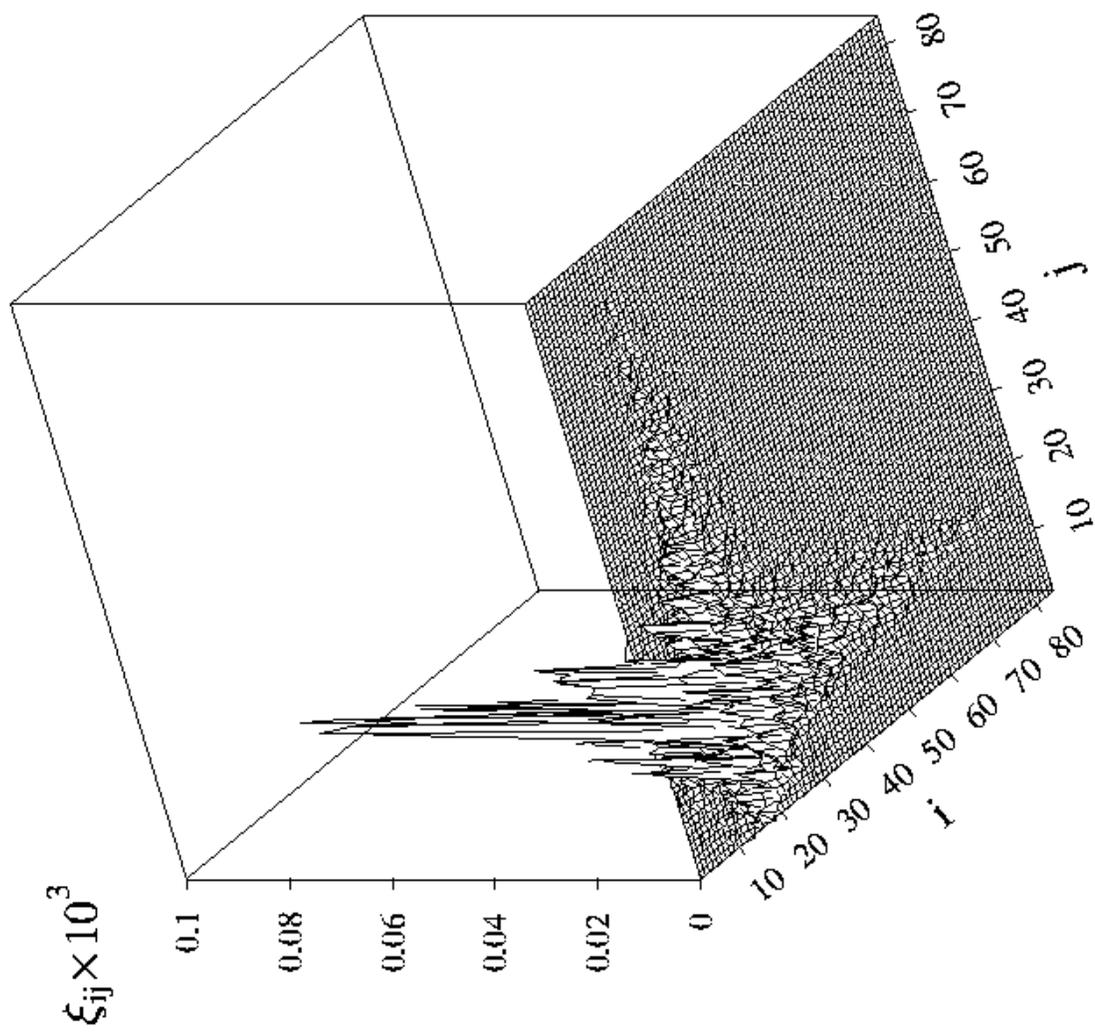
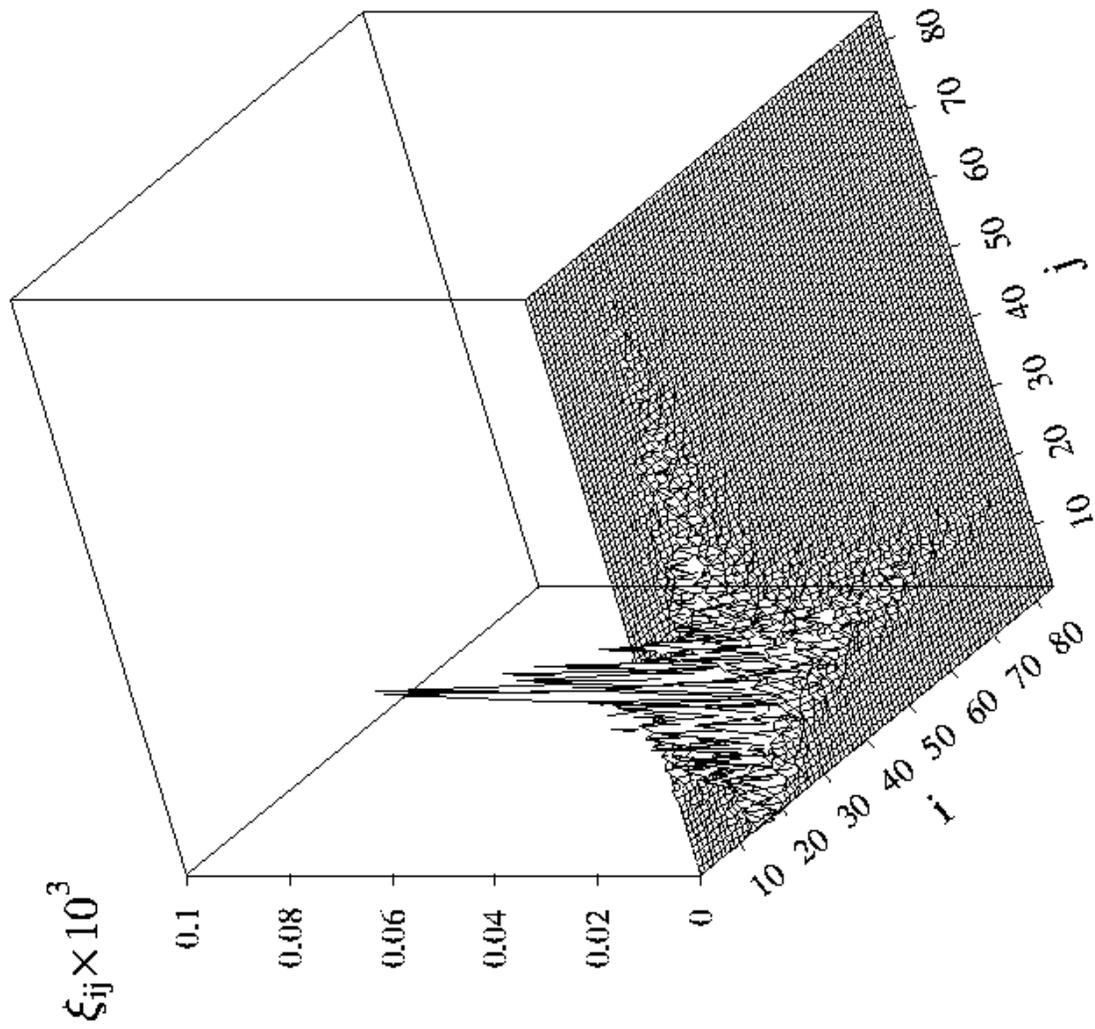
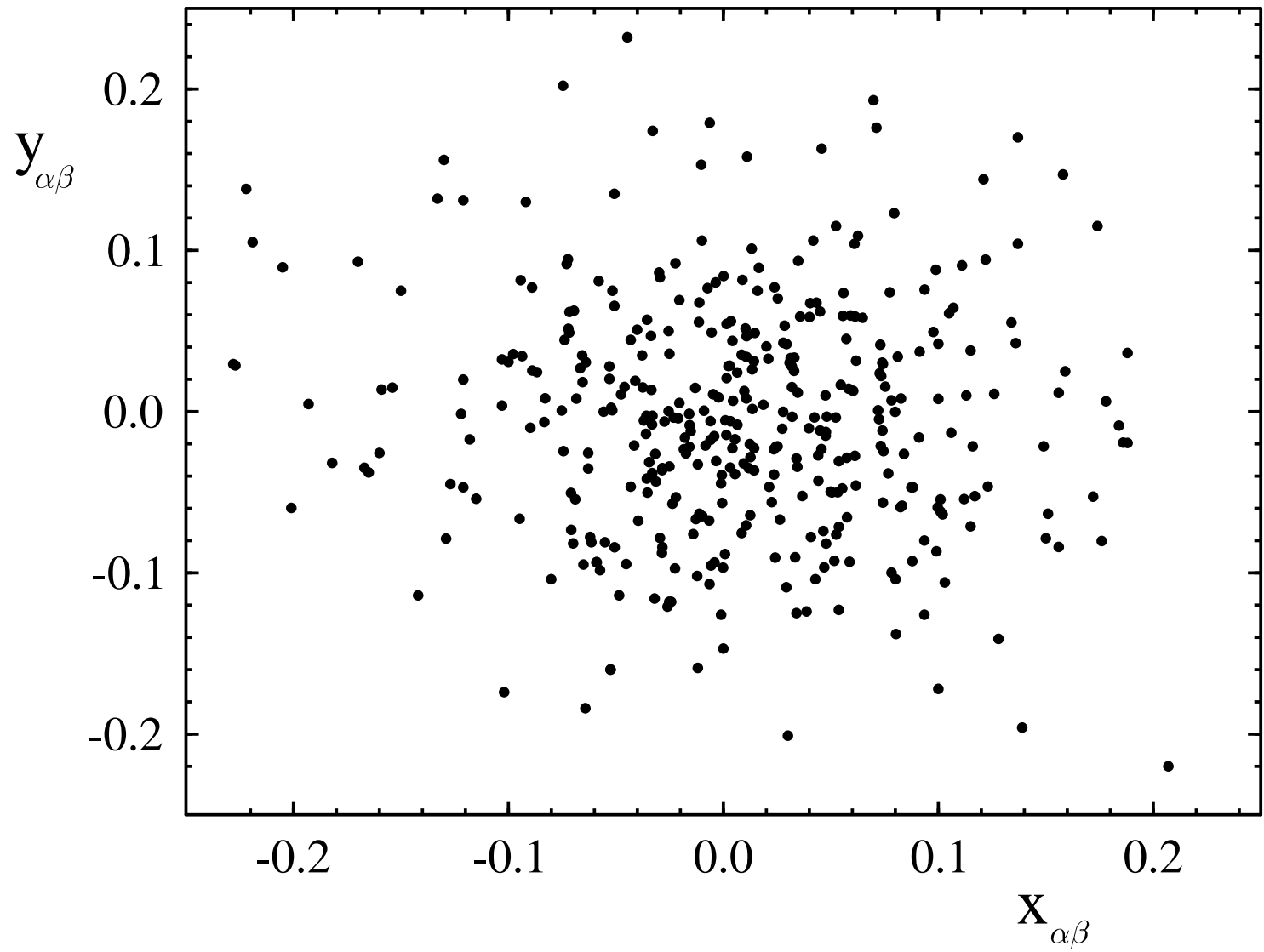
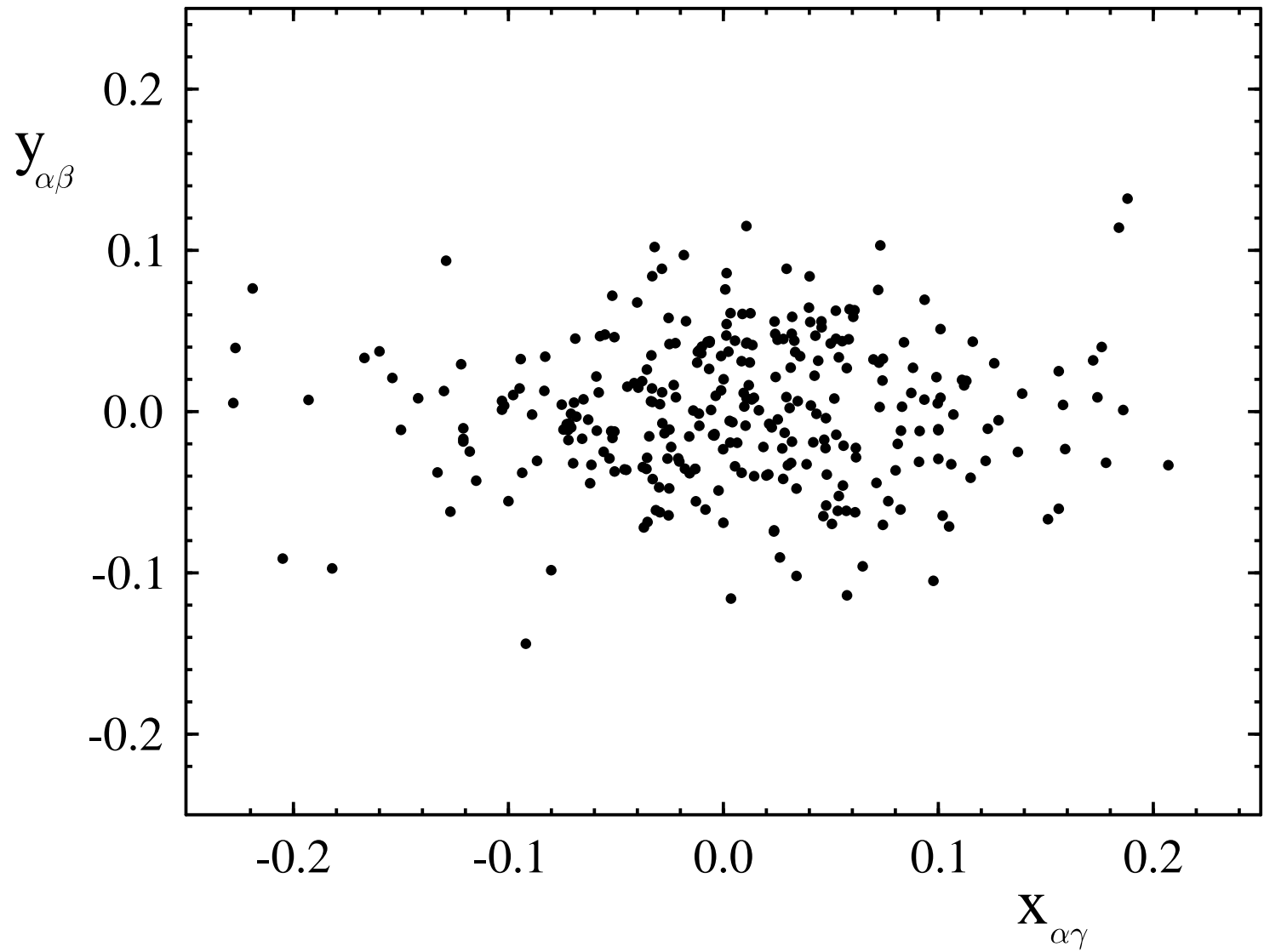
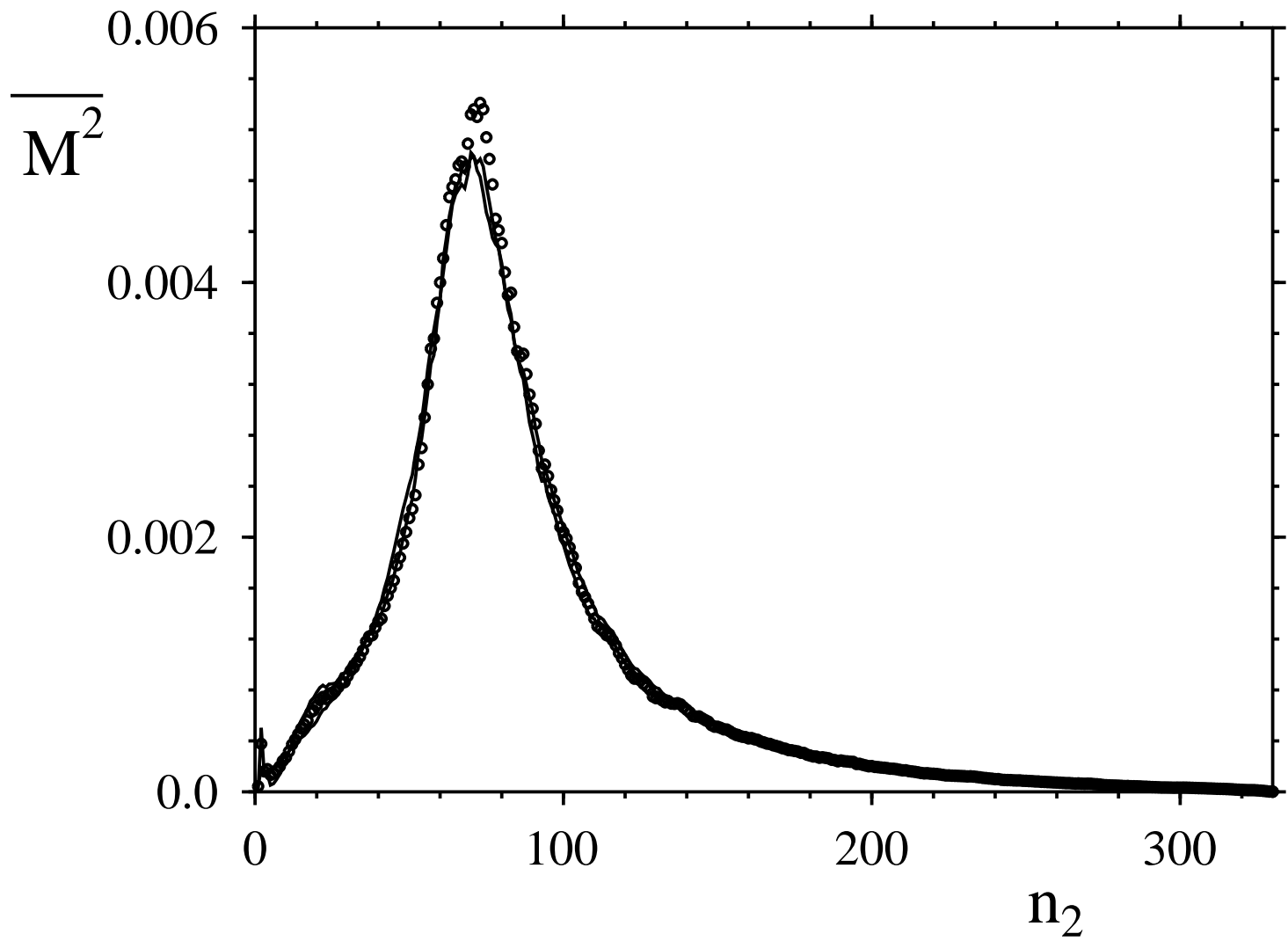
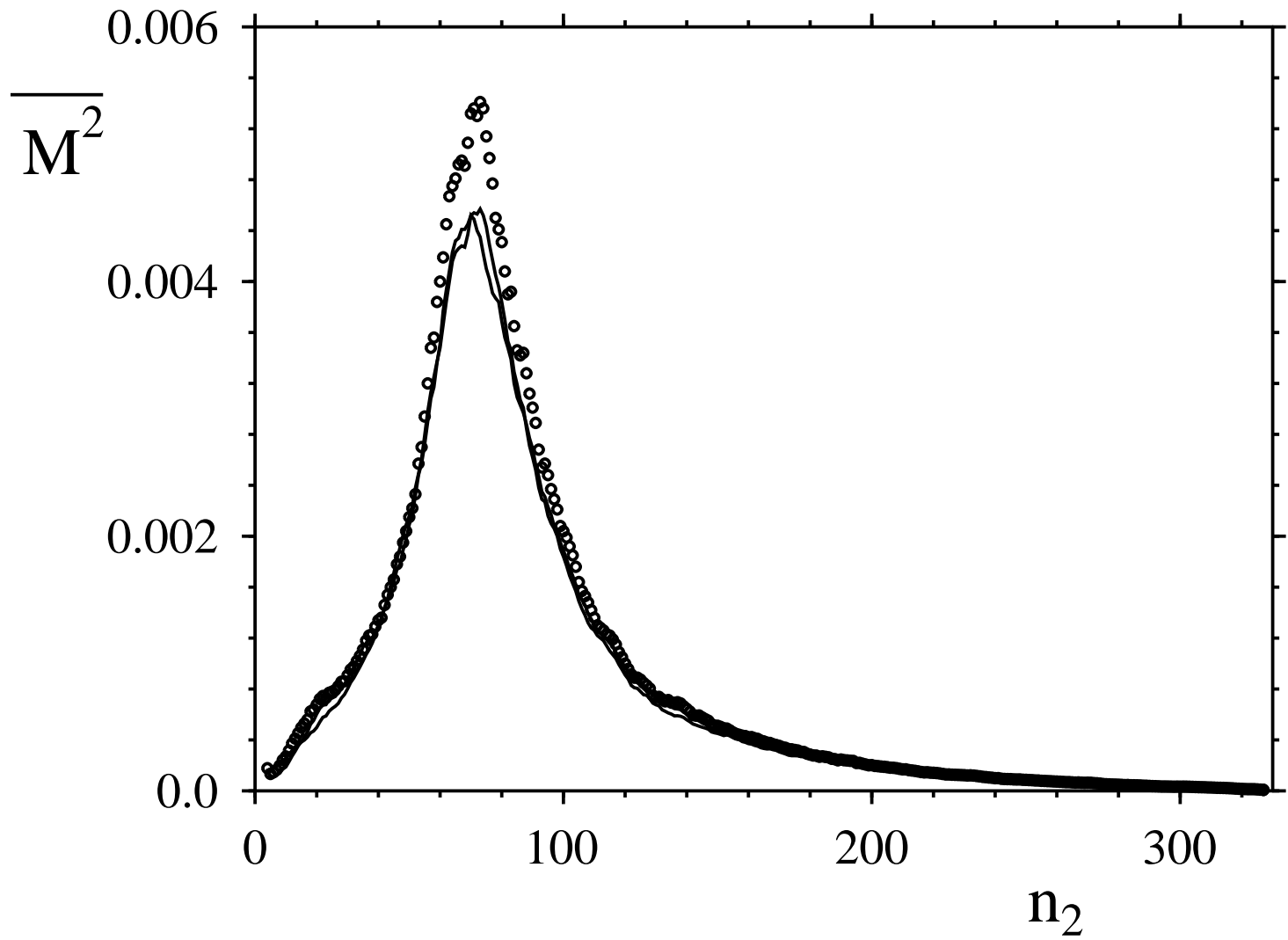
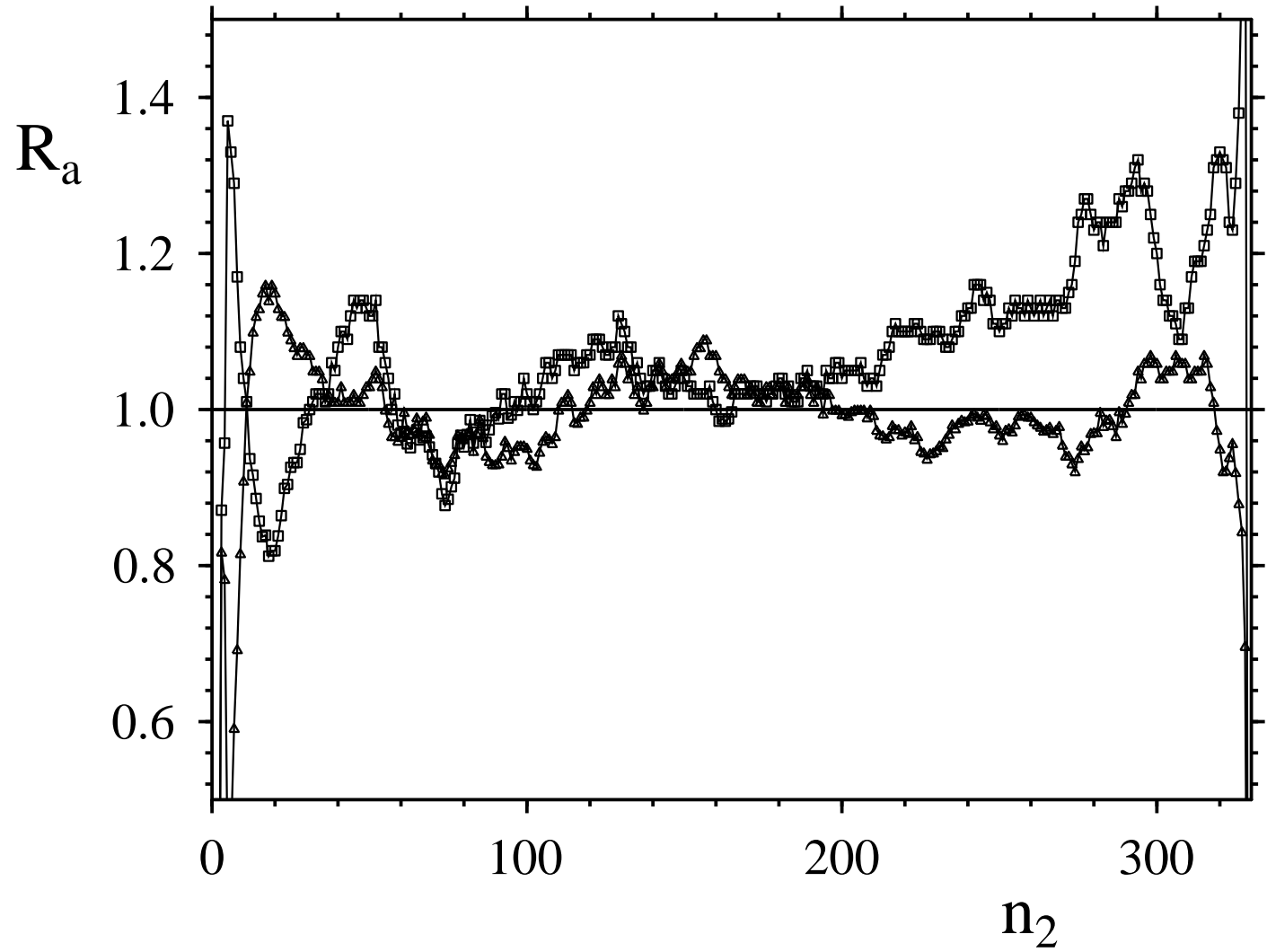
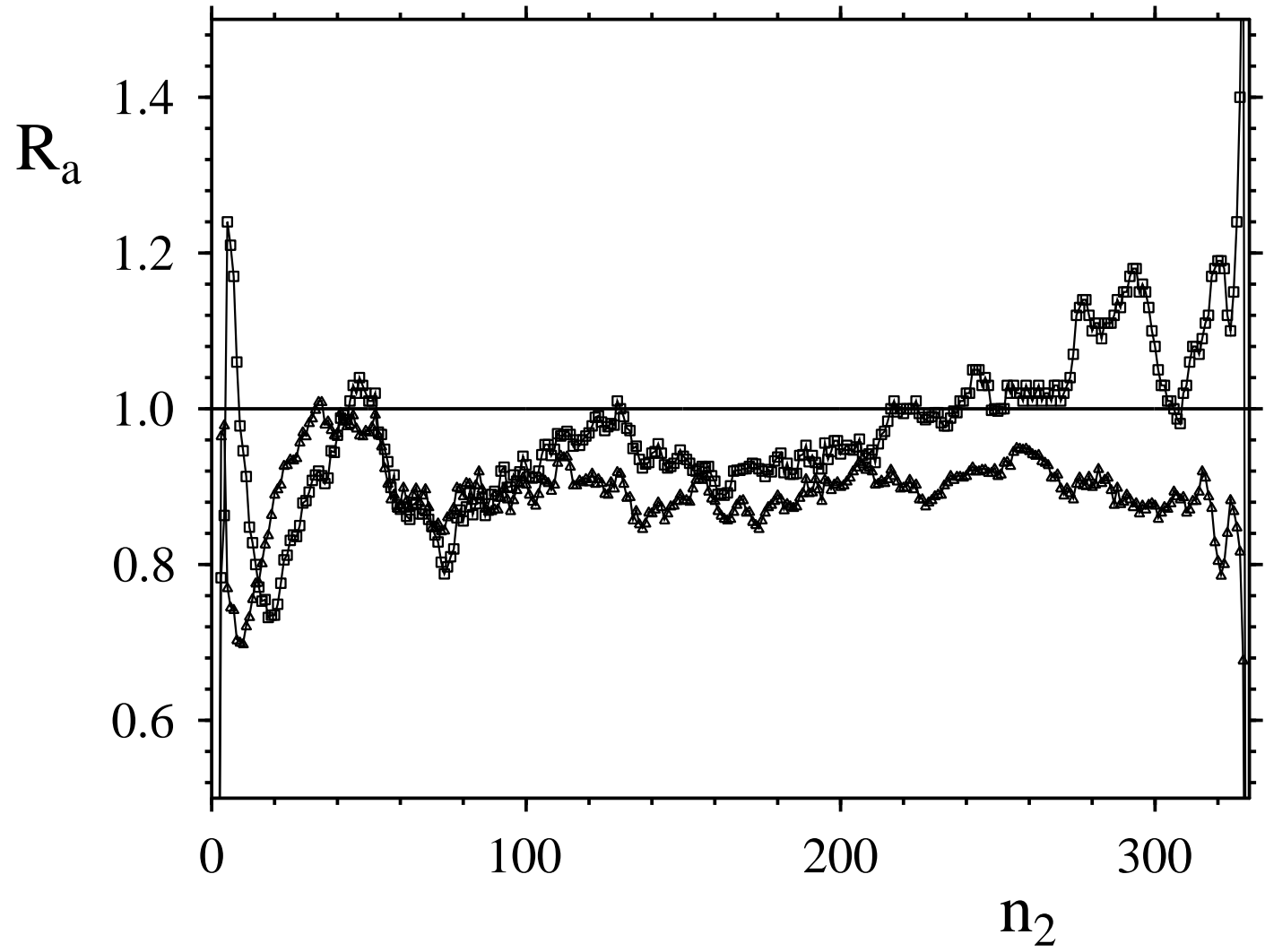
\end{document}